\documentclass[usenatbib,fleqn]{mnras}

\usepackage{newtxtext,newtxmath}

\usepackage[T1]{fontenc}
\usepackage{ae,aecompl}

\usepackage{graphicx}	
\usepackage{amsmath}	
\usepackage{comment}
\usepackage{tablefootnote}
\usepackage{pdflscape}	
\usepackage[htt]{hyphenat} 
\usepackage{color}
\hypersetup{
  colorlinks   = true, 
  urlcolor     = magenta, 
  linkcolor    = blue, 
  citecolor   = cyan 
}

\title[Colossal over- \& under-dense structures at $z<1$]
{Subaru Hyper Suprime-Cam excavates colossal over- and under-dense structures over 360 deg$^2$ out to  z = 1}

\author[R. Shimakawa et al.]
{Rhythm Shimakawa,$^{1}$\thanks{E-mail: rhythm.shimakawa@nao.ac.jp (RS)}
Yuichi Higuchi,$^{1,2}$
Masato Shirasaki,$^{1,3}$
Masayuki Tanaka,$^{1}$
\newauthor
Yen-Ting Lin,$^{4}$
Masao Hayashi,$^{1}$
Rieko Momose,$^{5}$
Chien-Hsiu Lee,$^{6}$
\newauthor
Haruka Kusakabe,$^{7}$
Tadayuki Kodama,$^{8}$
Naoaki Yamamoto,$^{8}$
\\
$^{1}$National Astronomical Observatory of Japan (NAOJ), National Institutes of Natural Sciences, Osawa, Mitaka, Tokyo 181-8588, Japan\\
$^{2}$Faculty of Science and Engineering, Kindai University, Higashi-Osaka, Osaka, 577-8502, Japan\\
$^{3}$The Institute of Statistical Mathematics, Tachikawa, Tokyo 190-8562, Japan\\
$^{4}$Academia Sinica Institute of Astronomy and Astrophysics, PO Box 23-141, Taipei 10617, Taiwan\\
$^{5}$Department of Astronomy, School of Science, The University of Tokyo, 7-3-1 Hongo, Bunkyo-ku, Tokyo 113-0033, Japan\\
$^{6}$NSF's National Optical-Infrared Astronomy Research Laboratory, USA\\
$^{7}$Observatoire de Gen\`{e}ve, Universit\'e de Gen\`{e}ve, 51 chemin de P\'egase, 1290 Versoix, Switzerland\\
$^{8}$Astronomical Institute, Tohoku University, 6-3, Aramaki, Aoba, Sendai, Miyagi, 980-8578, Japan
}

\date{Accepted XXX. Received YYY; in original form 2020 December 20}

\pubyear{2020}

\begin{document}
\label{firstpage}
\pagerange{\pageref{firstpage}--\pageref{lastpage}}
\maketitle

\begin{abstract}
Subaru Strategic Program with the Hyper-Suprime Cam 
(HSC-SSP) has proven to be successful 
with its extremely-wide area coverage in past years. Taking advantages of this 
feature, we report initial results from exploration and research of expansive over- 
and under-dense structures at $z=$ 0.3 -- 1 based on the second Public Data Release 
where optical 5-band photometric data for $\sim$ eight million sources with $i<23$ 
mag are available over $\sim360$ square degrees. We not only confirm known 
superclusters but also find candidates of titanic over- and under-dense regions 
out to $z=1$. 
The mock data analysis suggests that the density peaks would involve one or more 
massive dark matter haloes ($>10^{14}$ M$_\odot$) of the redshift, and the density troughs tend to be 
empty of massive haloes over $>10$ comoving Mpc. 
Besides, the density peaks and troughs at $z\lesssim0.6$ are in part identified 
as positive and negative weak lensing signals respectively, in mean tangential 
shear profiles, showing a good agreement with those inferred from the full-sky 
weak lensing simulation.
The coming extensive spectroscopic surveys will be able to resolve these colossal 
structures in three-dimensional space. The number density information over the 
entire survey field is available as grid-point data 
on the website of the HSC-SSP 
data release (\url{https://hsc.mtk.nao.ac.jp/ssp/data-release/}).
\end{abstract}




\section{Introduction}
\label{s1}

Mapping large-scale structures on cosmological scales and studying properties of 
galaxies therein across the cosmic time inform us of the role of large-scale 
environment on galaxy evolution in the hierarchical universe. In the local 
universe, more than a dozen studies and discussions have been made on the environmental 
dependence of galaxy evolution over such an extensive survey volume
\citep{Tanaka2004,Baldry2006,Sorrentino2006,Park2007,Peng2010,Alpaslan2015,Douglass2017}, 
based on data such as the Sloan Digital Sky Survey 
\citep{York2000,Eisenstein2011,Blanton2017}, the 2dF Galaxy Redshift Survey 
\citep{Colless2001}, 
and the Galaxy And Mass Assembly survey \citep{Driver2009,Driver2011,Liske2015}. 
Beyond the local universe, on the other hand, such efforts remain far from 
complete due to the lack of deep and extremely wide-field data.

Meanwhile, statistics of cosmic voids have been gaining a lot of attention in constraining 
cosmological parameters such as density contrast, dark energy parameter, 
and neutrino mass \citep{Regos1991,Dekel1994,Agarwal2011,Pisani2015,Contarini2019}, 
and to test modified gravity \citep{Clampitt2013,Cai2015}. Supervoids are 
considered to be a causal factor of a controversial feature seen in the cosmic 
microwave background, namely the so-called "cold spot" 
\citep{Inoue2006,Inoue2007,Szapudi2015,Finelli2016}. 
Such diffuse environments also provide us with galaxy growth driven by in-situ 
process \citep{Peebles2001,Rojas2005a}. However, it has been challenging to 
exhaustively investigate void galaxies beyond the local universe until recently 
as such an exploration is too expensive without wide-field imager or 
spectrograph on large aperture telescopes.

The circumstances prevailing today are no longer the same as in the past, thanks 
to the advent of wide field imaging surveys over $>100$ square (sq.) deg area, 
such as the Canada–France–Hawaii Telescope Lensing Survey \citep{Heymans2012}, 
the Kilo-Degree Survey \citep{DeJong2013}, the Dark Energy Survey (DES) 
\citep{Abbott2018}, and the Subaru Strategic Program with the Hyper Suprime-Cam 
\citep{Aihara2018}. 
The two-point correlation and power spectrum analyses of cosmic shear have 
proven successful as a powerful cosmological probe (e.g., 
\citealt{Heymans2013,Hildebrandt2017a,Hikage2019}).
More recently, \citet{Gruen2018} demonstrate cosmological constraints from the 
density contrast over the wide-field area by measuring galaxy number densities and 
gravitational lens shears based on DES First Year and SDSS data. 
Nevertheless, a number of scientific questions related to galaxy evolution still 
remain unanswered; the millions of galaxies from the aforementioned surveys thus 
provide us with wonderful data for more detailed investigations. 

Subaru Strategic Program (SSP) with the Hyper Suprime-Cam (HSC), the prime focus 
camera on the Subaru Telescope with an exceptionally wide-field view (1.5 deg in diameter) compared to 
other 8-meter class telescopes began in 2014
\citep{Miyazaki2018,Furusawa2018,Kawanomoto2018,Komiyama2018}. The program has now 
completed $\sim90$ per cent of the allocated observing runs, reaching $\sim$1,000 
sq. degrees as of this writing. HSC-SSP is designed to address a various range of 
fundamental questions such as weak-lensing cosmology and galaxy evolution out to 
the early Universe\footnote{\url{https://hsc.mtk.nao.ac.jp/ssp/science/}}. 
Such in-depth, wide-field data enable systematic explorations and surveys of 
unique large-scale systems out to high redshifts (see, e.g. a red-sequence cluster 
search at $z=$ 0.1 -- 1.1 by \citealt{Oguri2018}; a protocluster search at $z\sim4$ 
by \citealt{Toshikawa2018}). The high quality and large-volume data-set taken from 
the 8-meter class telescope permits statistical research of galaxy properties such 
as mass assembly histories and red fractions of galaxies in galaxy clusters 
\citep{Lin2017a,Nishizawa2018} and a morphological classification 
\citep{Tadaki2020} for more than ten thousands or millions of sources.

We thus started working on density mapping at $z>0.3$ on the cosmological scale to 
discover the very-large over- and under-dense structures like 
superclusters
\footnote{Generally, in this paper, massive clusters hosting 
$\geq10^{15}$ M$_\odot$ halo masses, or clustering clusters that may grow into 
such very massive clusters at $z=0$ are referred as superclusters. We also refer 
to under-dense regions over $>100$ comoving Mpc as supervoids.}
and cosmic voids 
\citep{DeVaucouleurs1953,Gregory1978,Kirshner1981,Oort1983}, and also on 
comprehensive analyses of galaxy characteristics as a function of the large-scale 
environments across the cosmic time. 
This paper reports initial results from our systematic search for gigantic over- 
and under-dense structures with HSC-SSP. As the first in a series of papers 
stemmed from our research, we focus on the density mapping over $\sim360$ square 
(sq.) degrees where the data are publicly accessible, which is less than one-third 
of the $\sim1,400$ sq. degrees to be covered at the end of the program. 
We overview the data-set in \S\ref{s2}, and then, explain the density estimation 
and present resultant maps (\S\ref{s3}). The discussion part (\S\ref{s4}) evaluates 
total masses associated with discovered colossal over- and under- densities based 
on a mock galaxy catalogue from a sizeable cosmological simulation and a weak 
lensing analysis. 
Lastly, we summarise highlights of this work and remark about the outlook for 
upcoming related programs (\S\ref{s5}). Throughout the paper, we adopt the AB 
magnitude system \citep{Oke1983}, and cosmological parameters of $\Omega_M=0.279$, 
$\Omega_\Lambda=0.721$, and $h=0.7$ in a flat Lambda cold dark matter model that 
are consistent with those from the WMAP nine year data \citep{Hinshaw2013}.
When we refer to figures or tables shown in this paper, we designate their initials 
by capital letters (e.g., Fig.~1 or Table~1) to avoid confusion with those in the 
literature.


\section{HSC-SSP catalogue}
\label{s2}

\begin{figure}
    \centering
	\includegraphics[width=.95\columnwidth]{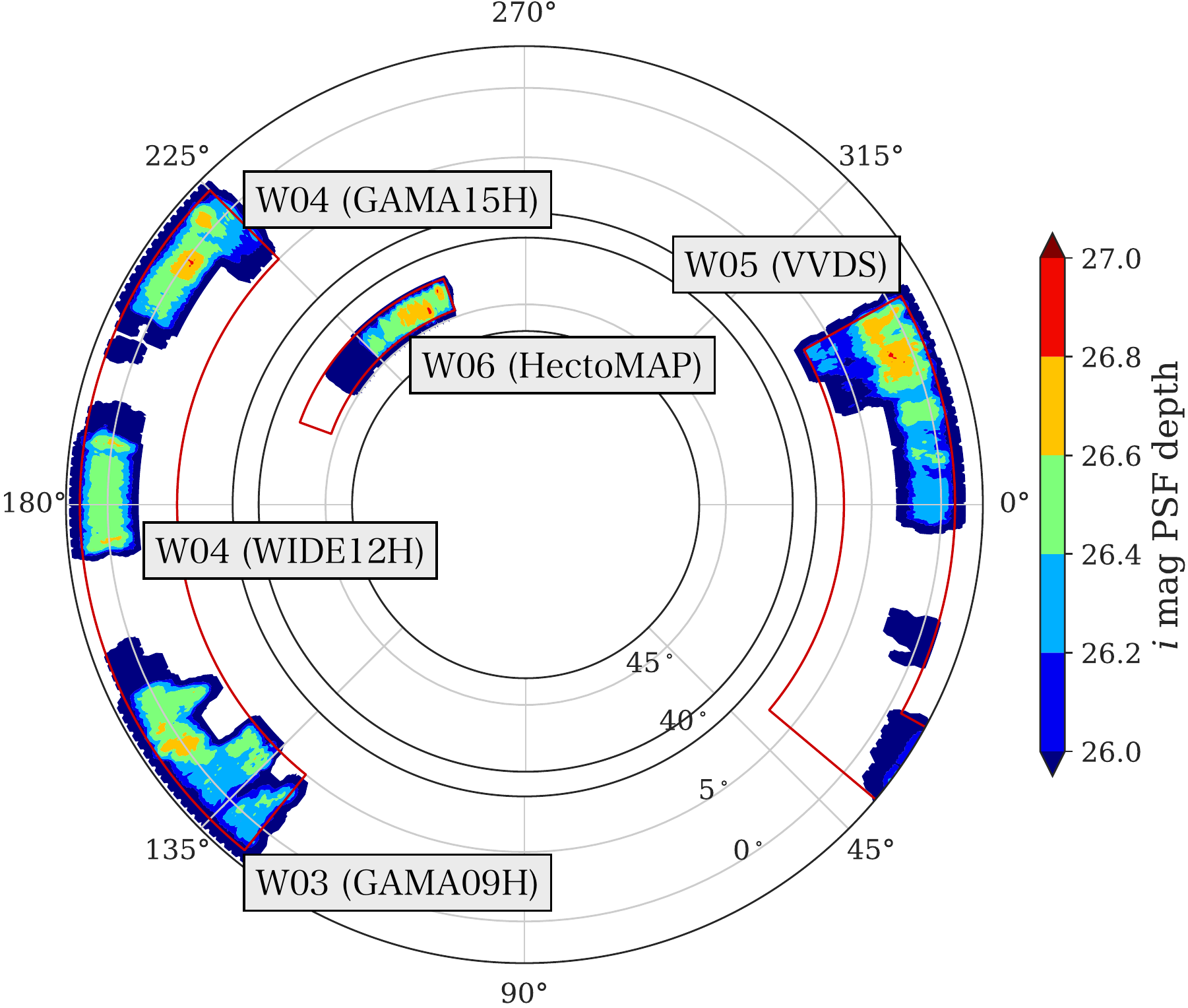}
    \caption{
    The above shows the sky coverage in PDR2 where colour-code means five sigma PSF 
    limiting magnitude in two arcsec diameter at $i$-band. We cut out five areas as 
    denoted on the sky map. The red rectangle area indicates the survey footprint 
    of HSC-SSP \citep{Aihara2018}.
    }
    \label{fig1}
\end{figure}

This work is based on the second Public Data Release (PDR2) of HSC-SSP, which 
became public on 2019 May 30 \citep{Aihara2019}. The HSC-SSP PDR2 contains the data 
taken during March 2014 and January 2018. The detailed information about PDR2 is 
summarised by \citet[table 2]{Aihara2019}. The database provides us with the 
science-ready catalogue through the data processing by the dedicated pipeline 
called {\tt hscPipe} \citep{Bosch2018}. We first collect bright sources in the 
$i$-band ($<23$ mag) with more than five sigma detection in {\tt cmodel} 
measurement in all HSC bands at 0.4 -- 1.1 $\mu$m ($g/r/i/z/y$). 
As for the {\tt cmodel} measurements, we refer readers to \citet[\S4.9.9]{Bosch2018}. 
We here exclude sources near bright stars ($<17.5$ mag), and those affected by 
cosmic rays, bad pixels, or saturated pixels using the following flags on the 
HSC-SSP database: 
{\tt pixelflags\_edge}, {\tt pixelflags\_interpolatedcenter}, 
{\tt pixelflags\_saturatedcenter}, {\tt pixelflags\_crcenter}, 
{\tt pixelflags\_bad}, {\tt pixelflags\_bright\_objectcenter}, and 
{\tt mask\_s18a\_bright\_objectcenter}
(\citealt{Coupon2018,Aihara2019}; \citealt[table 2]{Bosch2018}).

\begin{figure}
    \centering
	\includegraphics[width=.9\columnwidth]{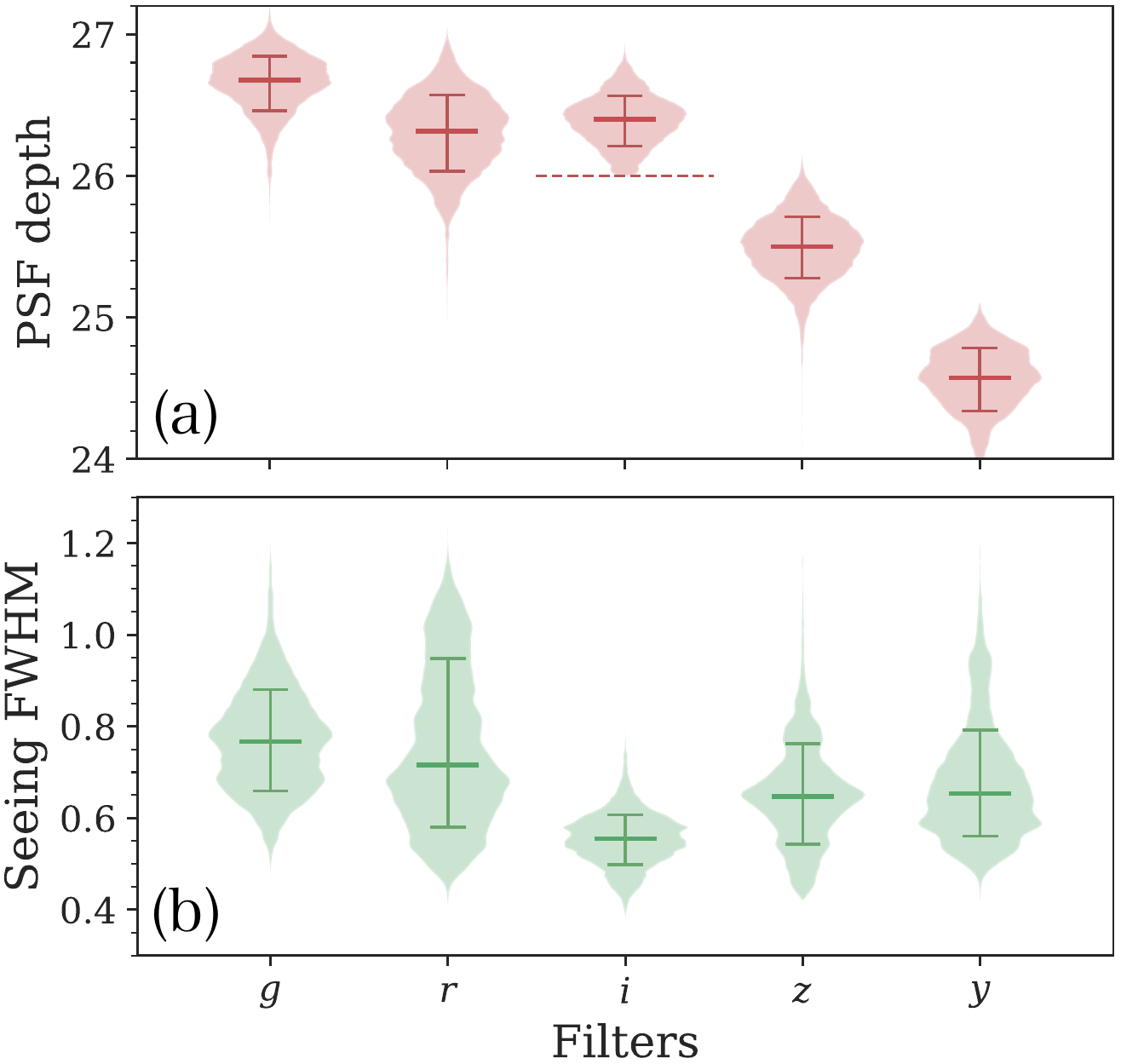}
    \caption{
    Violin plots of (a) the image depth and (b) the seeing size distributions in 
    $g/r/i/z/y$ broadband filters of HSC. The width of each violin is scaled based 
    on the Gaussian kernel density estimation (KDE). We employ the survey field in 
    which the PSF limiting magnitude in $i$-band is deeper than 26 mag (the dashed 
    line).
    The horizontal lines in each distribution show the median value and the 
    intervals of 68th percentile.
    }
    \label{fig2}
\end{figure}

We employ an area totally $\sim360$ sq. degrees from the available HSC-SSP Wide 
layer (\citealt[figure 1]{Aihara2019}; see also Fig.~\ref{fig1}) in which five sigma limiting magnitude of 
point-spread-function (PSF) in two arcsec diameter is deeper than 26 mag in 
$i$-band. These criteria are chosen to prevent the need for correction of the 
variance of number densities due to depth variation across the survey field 
(see Fig.~\ref{fig2}, \ref{fig3} and a detailed description in \S\ref{s31}). One 
should note that we exclude two fields, W01 (WIDE01H) and W02 (XMM) regions 
\citep{Aihara2019}, which 
cover much narrower areas than the other fields (W03--W06) in the PDR2 database. After that, 
we select samples in a photometric redshift (photo-$z$) range from $z=$ 0.3 to 1 
using {\tt Mizuki}, an SED-based photo-$z$ code \citep{Tanaka2015}. 
The biweight dispersion of $\Delta z=|z_\mathrm{spec}-z_\mathrm{photo}|$ and the 
outlier rate ($|\Delta z|>0.15$) for the HSC sample at $i<23$ mag are $\leq0.04$ 
and $\sim0.1$, respectively \citep[table 2]{Tanaka2018}. We only employ objects 
with the reduced chi-square $\chi_\nu<5$ of the best-fitting model following the 
suggestion in the literature \citep[\S7]{Tanaka2018}. 
We carry out a further selection cut for the redshift $z=$ 0.4 -- 0.5 bin where 
there is a strong Balmer--Lyman break degeneracy. Since misclassified objects at 
$z\sim3$ tend to show very young ages ($\sim1$ Gyr) with high specific 
star-formation rates (SSFRs), we omit such samples which have SSFRs higher than 
1 Gyr$^{-1}$ to minimise the potential contaminants of Lyman break galaxies at 
$z\sim3$. We note that the contamination of Lyman break galaxies for higher 
redshift bins, such as $i$-band dropouts, is negligible given our detection 
criteria in all HSC bands, including $g$-band. Through these criteria, our 
galaxy sample contains about ten million objects. 

\section{Mapping over- and under-densities}
\label{s3}

\subsection{Density estimation}
\label{s31}

We count the galaxy number within a top-hat aperture of 
$r=$ 10 and 30 arcmin ($n_{r=10'}$ and $n_{r=30'}$) at seven projected redshift 
slices: $z=$ [0.3:0.4), [0.4:0.5), [0.5:0.6), [0.6:0.7), [0.7:0.8), [0.8:0.9), 
and [0.9:1.0). A spatial resolution of the projected density map is set to be 
about 1.5 arcmin that is sufficiently smaller than the aperture sizes. 
The aperture radius corresponds to 4.0, 5.0, 6.0, 6.9, 7.8, 8.6, 9.4 comoving Mpc 
(cMpc) in $r=10$ arcmin and 12.1, 15.1, 18.0, 20.8, 23.4, 25.8, 28.1 cMpc in 
$r=30$ arcmin at $z=$ 0.35, 0.45, 0.55, 0.65, 0.75, 0.85, 0.95, respectively. As a 
benchmark study of the series of papers, we choose these fixed aperture radii by 
referencing to those typically used in the Dark Energy Survey 
\citep{Gruen2016,Gruen2018}. Since this work focuses on enormously large 
structures ($\gtrsim20$ cMpc) rather than local (sub-)structures, recent 
sophisticated density estimations (e.g., \citealt{Sousbie2011,Lemaux2017}) are not 
crucial. 
Furthermore, the choice of the fixed angular aperture is motivated to provide 
functional usability with catalogue users. Since our density map catalogue 
includes the number densities within the fixed apertures at narrower redshift bins 
($\Delta z=0.02$), users can reconstruct density maps at a specific redshift space 
between $z=0.3$ and 1 (Appendix \ref{a1}).
Besides, we prefer to use the fixed angular aperture method to keep consistency of 
a number density correction around holes by bright-star masks and boundaries 
across redshifts and fields as mentioned below. However, the density catalogue 
also contains physical aperture measurements of $r=10$ co-Mpc given the use for 
more scientific purpose, which provide complementary information to the fixed 
angular aperture estimations. 

\begin{figure}
    \centering
	\includegraphics[width=.9\columnwidth]{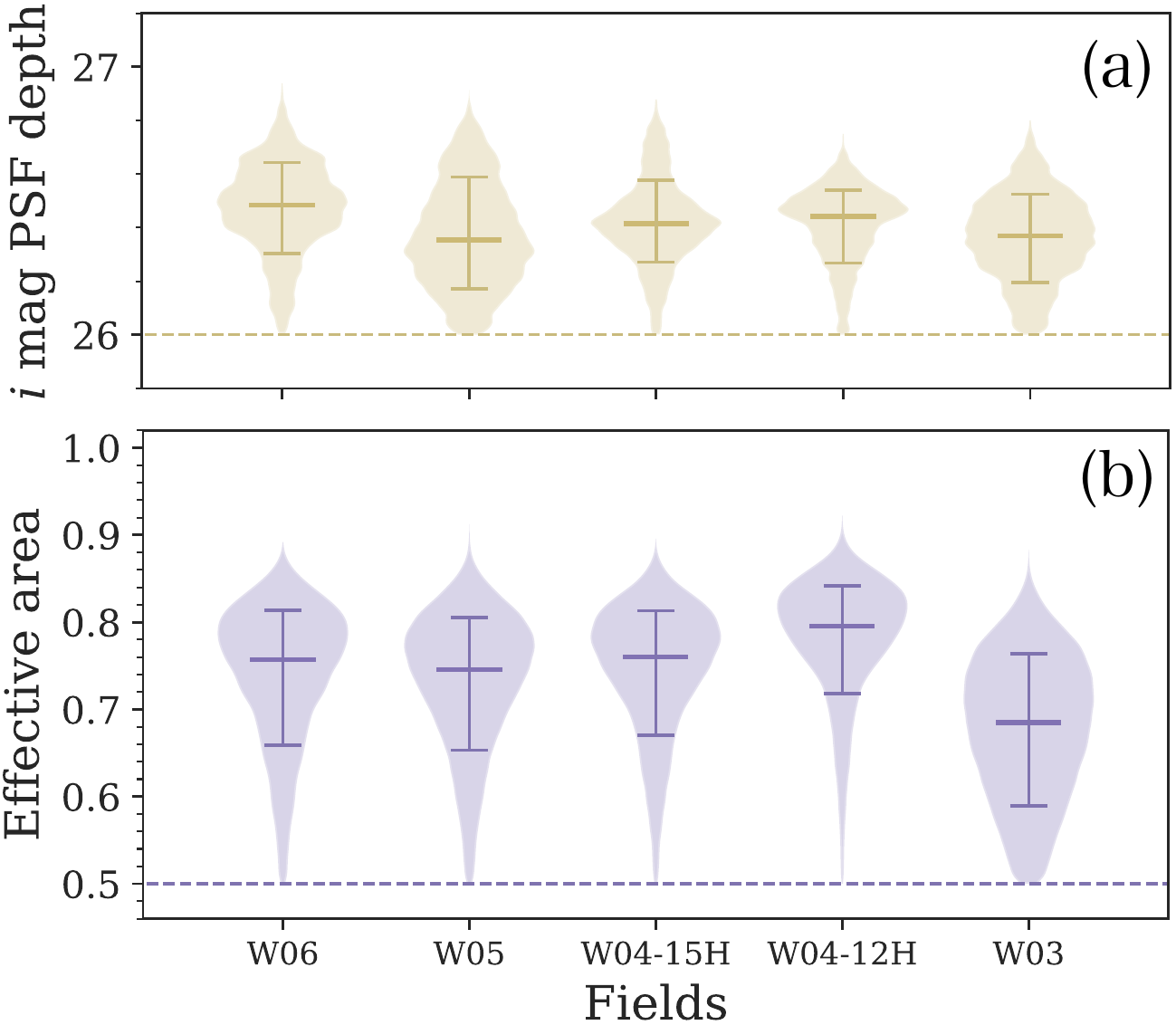}
    \caption{
    Violin plots for  the field variation of (a) the $i$-band limiting magnitude 
    and (b) the effective area, i.e., the fraction of unmasked area. As in 
    Figure~\ref{fig2}, the median and 68th percentile values are shown in each 
    bin. 
    }
    \label{fig3}
\end{figure}

\begin{figure*}
    \centering
	\includegraphics[width=.95\textwidth]{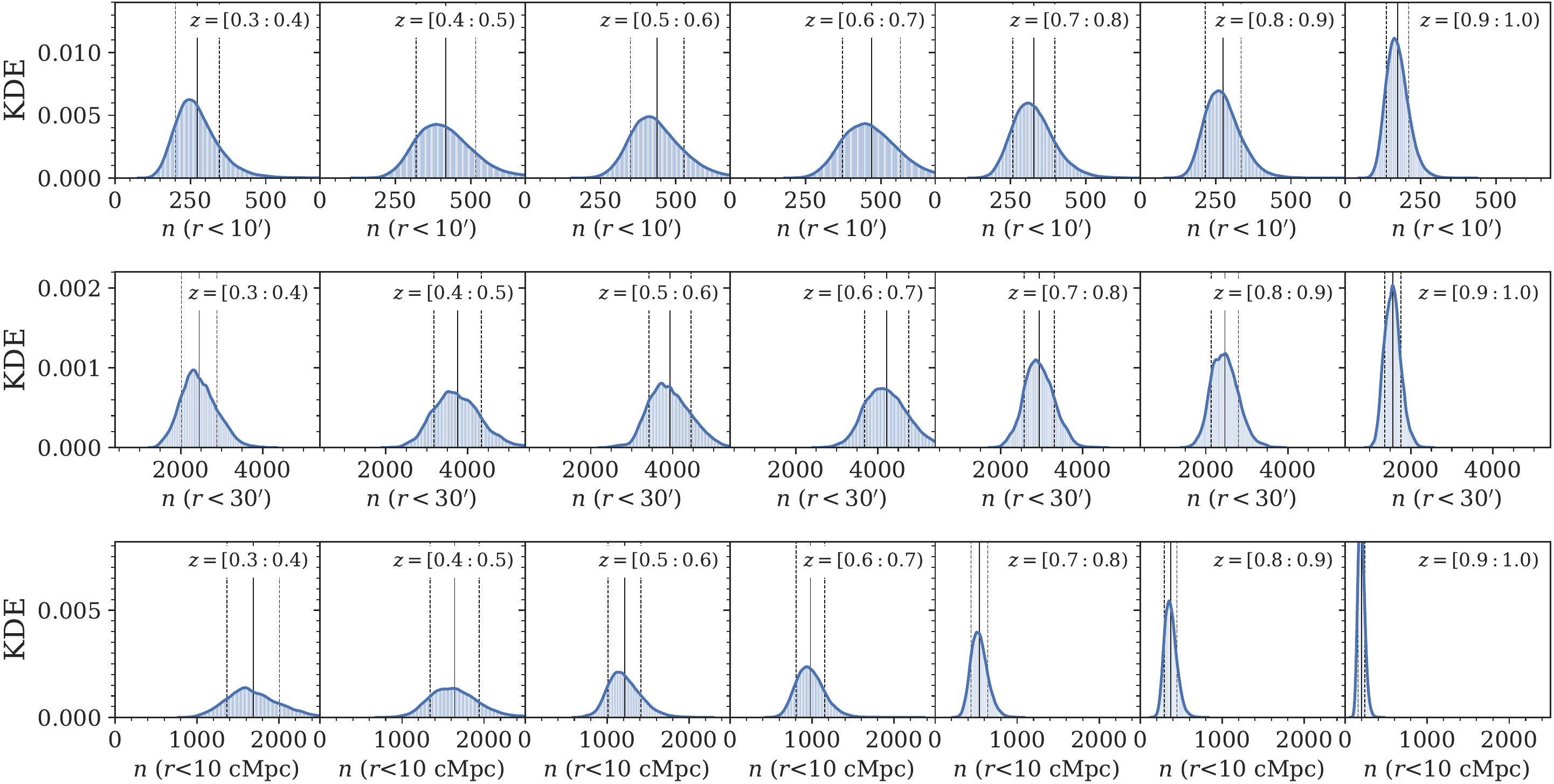}
    \caption{The number density distribution (per aperture) in Gaussian kernel 
    density estimation (KDE) at each redshift range from $z=0.3$ to 1. Those in 
    upper and lower panels respectively show the number of $i$-band magnitude 
    limited ($i<23$) sources within $r=10$ and $r=30$ arcmin apertures. 
    The black solid and dashed lines mean the average and $\pm1\sigma$ standard 
    deviations for each panel. Here the mask correction has been implemented.
    }
    \label{fig4}
\end{figure*}

Deriving proper number densities across the entire survey field must need 
appropriate bright-star masking to determine the effective survey area. For that 
reason, we implement the random catalogue provided by PDR2, which allows us to 
evaluate the fraction occupied by the bright-star masks and the boundaries of 
survey fields in each aperture area for the density estimate 
\citep{Coupon2018,Aihara2019}. This work employs only the survey areas that are 
masked less than 50 per cent at the corresponding grid point. 
We then apply the mask correction by dividing number densities by unmasked 
fraction (defined as effective area) to obtain the absolute number densities in 
each aperture area. One should note that we confirm no clear trend in the median 
values of mask-corrected number densities toward the effective area, meaning that 
the mask correction is not systematically over-estimated nor under-estimated. 
The derived number density distributions in the seven redshift slices with two 
aperture sizes are summarised in Figure~\ref{fig4}. The area of the five fields 
employed amounts to 359.5 sq. deg in $r=10$ arcmin aperture measurement (or 
261.2 sq. deg of the effective area used in the actual calculation; see 
Table~\ref{tab1}), which is the most extensive ever with such a deep optical 
multi-band photometry. The total number of $i$-band selected sources associated 
with the whole survey area reaches 7.8 million. Future papers will delve into 
those characteristics with respect to various cosmic environments. 

\begin{figure*}
    \centering
	\includegraphics[width=.8\textwidth]{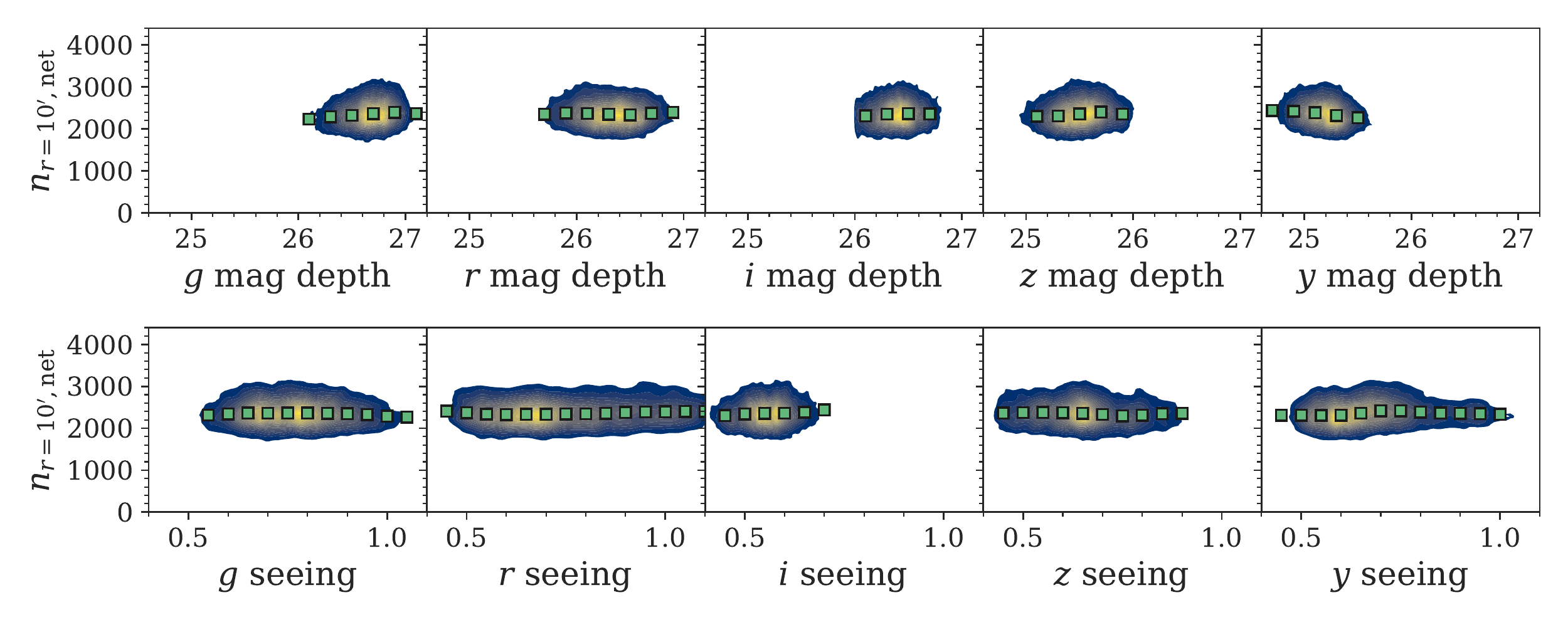}
    \caption{
    Distributions of the net number densities given by summing all number 
    densities in the seven redshift slices at $z=$ 0.3 -- 1 in each aperture, as 
    a function of five sigma limiting magnitude (upper panels) and seeing sizes 
    (lower panels) in $g/r/i/z/y$ bands. The green squares show the median values 
    within small bins of the depth ($\Delta=0.2$ mag) and the seeing size 
    ($\Delta=0.05$ arcsec).
    }
    \label{fig5}
\end{figure*}

As a sanity check, we see if the net number densities combined in all 
redshift bins from $z=0.3$ to 1 have any systematic bias depending on image depth 
or seeing size in each broadband filter (Fig.~\ref{fig2} and \ref{fig3}). We here 
adopt the number densities within $r=10$ arcmin apertures which is more comparable 
to the size of individual HSC patch ($\sim12\times12$ arcmin$^2$; see 
\citealt[\S4]{Aihara2018}) than that with $r=30$ arcmin apertures. Then, we 
confirm that they are constant relative to various image depths and seeing sizes 
within the margin of error of less than five per cent (Fig.~\ref{fig5}). 

\begin{table}
 \caption{Map information. The first and second columns are the field id and name. 
 The third and fourth columns show the survey area and the number of density grids 
 where we calculate the number count of targets, which include the outer edges of 
 the survey fields that we do not employ in the analyses. The spatial resolution 
 is about 1.5 arcmin. 
 These columns may be helpful for ones who use the 
 online catalogue (Appendix~\ref{a1}). 
 The fifth column indicates the actual survey area satisfying our data 
 requirements and the values in parentheses are the effective field coverage with 
 the effective area $>50$ per cent in $r=10$ arcmin apertures (which are 
 marginally larger in $r=30$ arcmin).}
 \label{tab1}
 \begin{tabular}{lllrr}
  \hline
  ID & Field & $\alpha_{min}$: $\alpha_{max}$ & N grids & Area deg$^2$ \\
        & & $\delta_{min}$: $\delta_{max}$ &         & (effective) \\
  \hline
  1 & W06 HectoMAP & $\alpha$=224.0:250.0 & 754$\times$120  & 38.2 \\
                 & & $\delta$=42.0:45.0   &    & (28.2) \\
  2 & W05 VVDS     & $\alpha=$330.0:363.0 & 1320$\times$200 & 109.4 \\
                 & & $\delta$=$-$1.0:4.0  &    & (80.0) \\
  3 & W04 GAMA15H  & $\alpha$=206.0:226.0 & 800$\times$160  & 56.6 \\
                 & & $\delta$=$-$2.0:2.0  &    & (42.2) \\
  4 & W04 GAMA12H  & $\alpha$=173.0:190.0 & 680$\times$160  & 52.4 \\
                 & & $\delta$=$-$2.0:2.0  &    & (40.8) \\
  5 & W03 GAMA09H  & $\alpha$=128.0:154.0 & 1040$\times$260 & 103.0 \\
                 & & $\delta$=$-$2.0:4.5  &    & (69.9) \\
  \hline
 \end{tabular}
\end{table}

\subsection{Projected density map}
\label{s32}

\begin{figure*}
    \centering
	\includegraphics[width=0.9\textwidth]{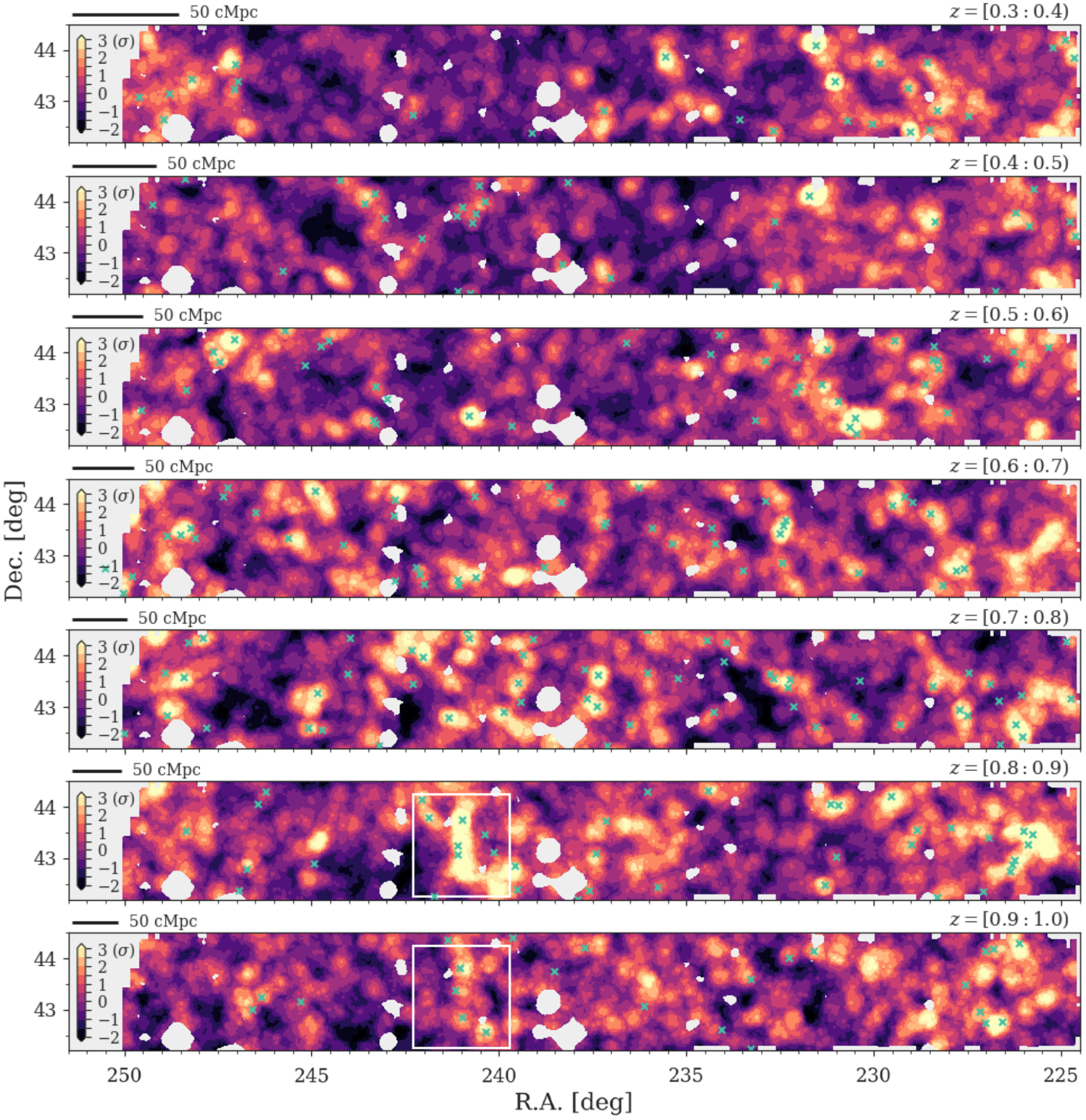}
    \caption{
    Density maps in W06 (HectoMAP) at $z=$ 0.3 -- 1 in step with $\Delta z=0.1$ 
    from top to bottom. 
    Colours indicate the number densities within $r=10$ arcmin in standard 
    deviation ($\sigma$=[$-2$: $+3$]) as denoted in left colour-bars. The eggshell 
    white area is the masked region with effective area of $\le50$ per cent. The 
    green crosses show the CAMIRA cluster samples with richness $\ge15$ at each 
    redshift (\citealt{Oguri2018} including the incremental update based on the 
    HSC-SSP PDR2). Density maps for the other fields are available as additional 
    figures through the online journal.
    The white boxed areas at $z=$ 0.8 -- 1.0 locate the whole structure 
    surrounding the CL1604 supercluster \citep{Hayashi2019}.
    }
    \label{fig6}
\end{figure*}

\begin{figure*}
    \centering
	\includegraphics[width=0.9\textwidth]{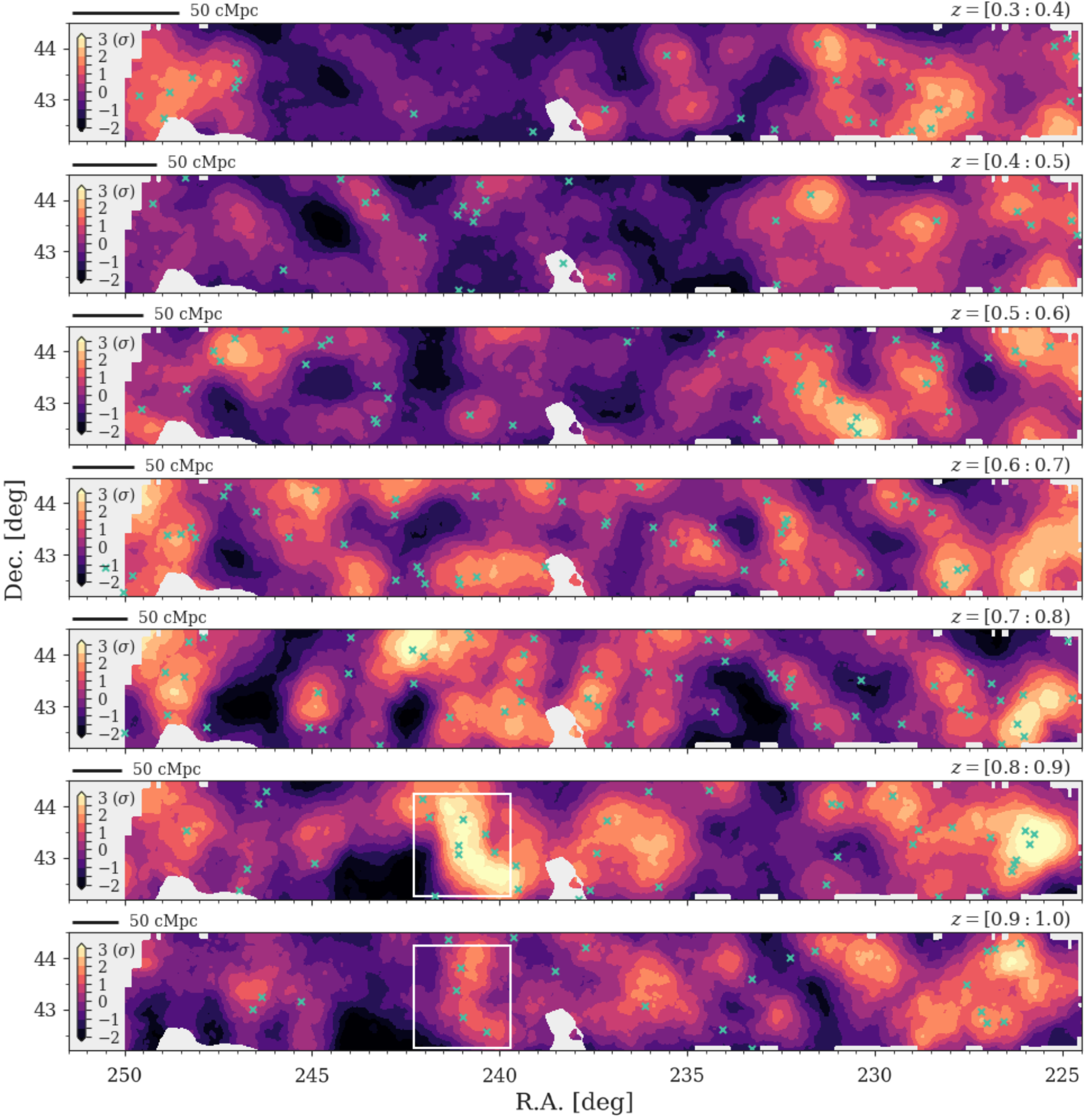}
    \caption{
    Same as Figure~\ref{fig6} but the number densities within $r=30$ arcmin.
    }
    \label{fig7}
\end{figure*}

\begin{figure*}
    \centering
	\includegraphics[width=0.9\textwidth]{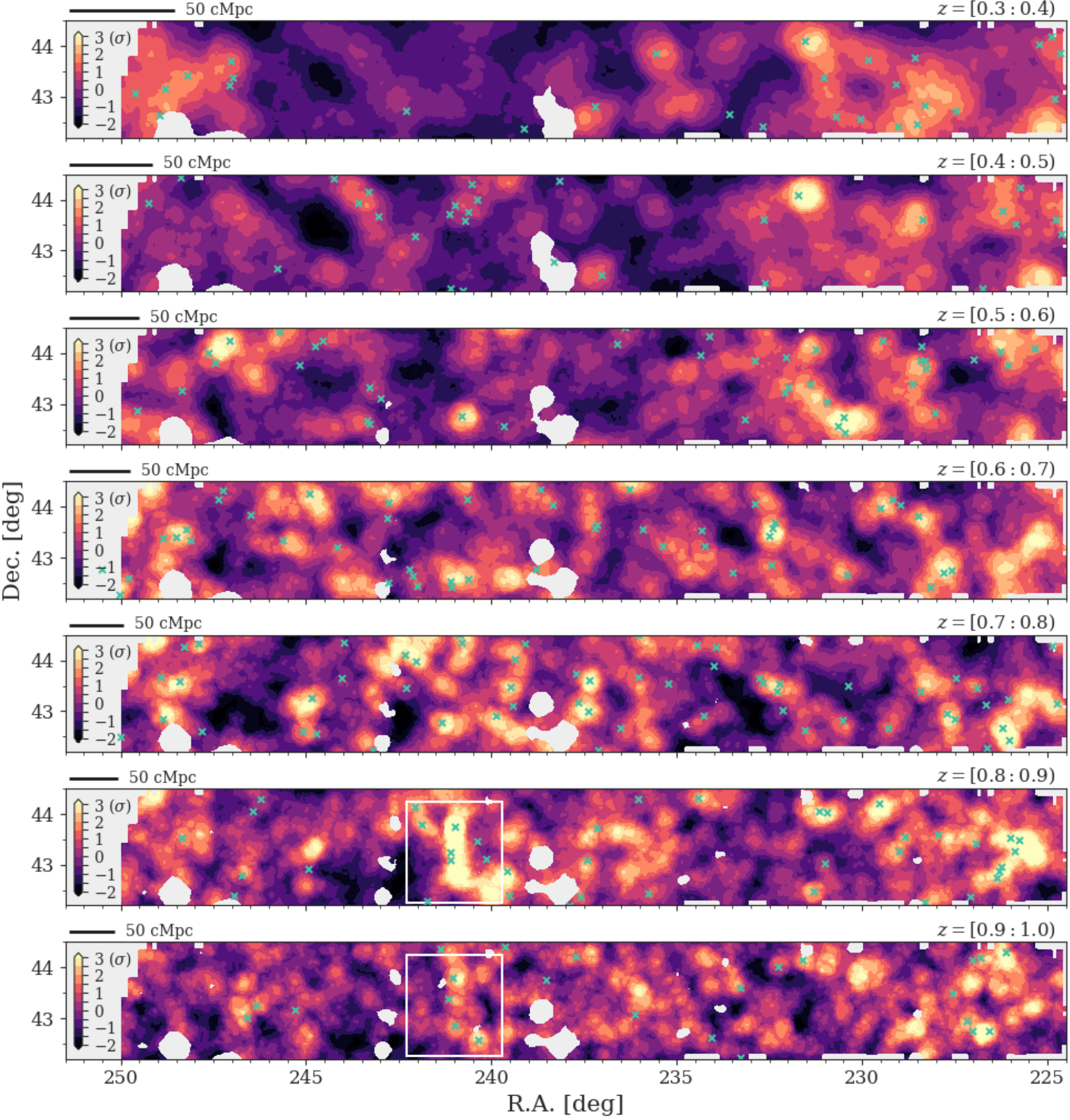}
    \caption{
    Same as Figure~\ref{fig6} but the number densities within $r=10$ cMpc.
    }
    \label{fig8}
\end{figure*}

Figure~\ref{fig6} and \ref{fig7} show the resultant density maps for one of five 
fields (W06), which largely overlaps with the HectoMAP region 
\citep{Geller2011,Geller2015}. The maps are colour-coded by the standard deviation 
($\sigma$) to the entire number density distribution (i.e., in all five survey 
fields) at each redshift (Fig.~\ref{fig4}). The area where a mask fraction is 
higher than 50 per cent is fully masked and is not taken into account in the 
following analyses. Density maps for the other fields can be found as additional 
figures through the online journal. Also, grid-point data of the density maps 
shown in this paper are available online (see Appendix~\ref{a1} for details). 

We mark positions of galaxy clusters identified by \citet{Oguri2018} based on 
the CAMIRA algorithm \citep{Oguri2014} to cross-check our density measurement with 
an existing catalogue.\footnote{include the incremental update based on HSC-SSP PDR2}  
One should note that our density estimate is largely 
different from their selection technique: \citet{Oguri2018} searched for galaxy 
clusters based on luminous red galaxies within a radius $\lesssim1$ physical Mpc 
at more flexible redshift spaces. On the other hand, we calculate number densities 
in much larger volumes at fixed redshift ranges. 
In general, none the less, the over-densities in Figure~\ref{fig6}-\ref{fig8} 
show a good agreement with distributions of the CAMIRA clusters. Interestingly, the CAMIRA clusters seem to 
trace the large-scale structures shown by the projected density maps of this work. 
For reference, 66 (or 36) and 47 (or 20) per cent of the CAMIRA clusters with 
richness $>15$ (see \citealt{Oguri2014} about the definition) are embedded in $>1$ 
sigma (or $>2$ sigma) densities of $r=10$ arcmin aperture estimation at $z=$ 0.3 
-- 0.4 and $z=$ 0.9 -- 1, respectively. More mismatch seen at higher redshifts 
would be mainly due to the larger physical aperture sizes at higher redshift bins 
in our density calculation since we confirm that the fractions increase when 
choosing smaller aperture sizes. It may also be affected by photo-$z$ 
uncertainties and/or fragmentation of massive systems in the earlier universe. 

A phenomenal structure seen at $z=$ 0.8 -- 0.9 ($\alpha=241^\circ$, 
$\delta=43^\circ$) is the CL1604 supercluster \citep{Gunn1986,Lubin2000,Gal2008}. 
A more panoramic picture of this supercluster reaching a 50 cMpc scale is recently 
reported by \citet{Hayashi2019}, which is consistent with what is seen in 
Fig.~\ref{fig6}-\ref{fig8}. 
We should note that the redshift projections at $z=$ [0.8:0.9) and [0.9:1.0) as 
shown in the figures are somewhat unsuitable for the visualisation of the CL1604 
supercluster since they split the whole redshift distribution of CL1604 centring 
on $z\sim0.9$ \citep{Gal2008}. 
Another massive structure appears on the west side at the same redshift, involving
six rich CAMIRA clusters. Because of these two enormous structures, the W06 field 
shows six per cent higher mean number density compared to that of the whole survey 
area at $z=$ 0.8 -- 0.9. 

\begin{figure*}
    \centering
	\includegraphics[width=.95\textwidth]{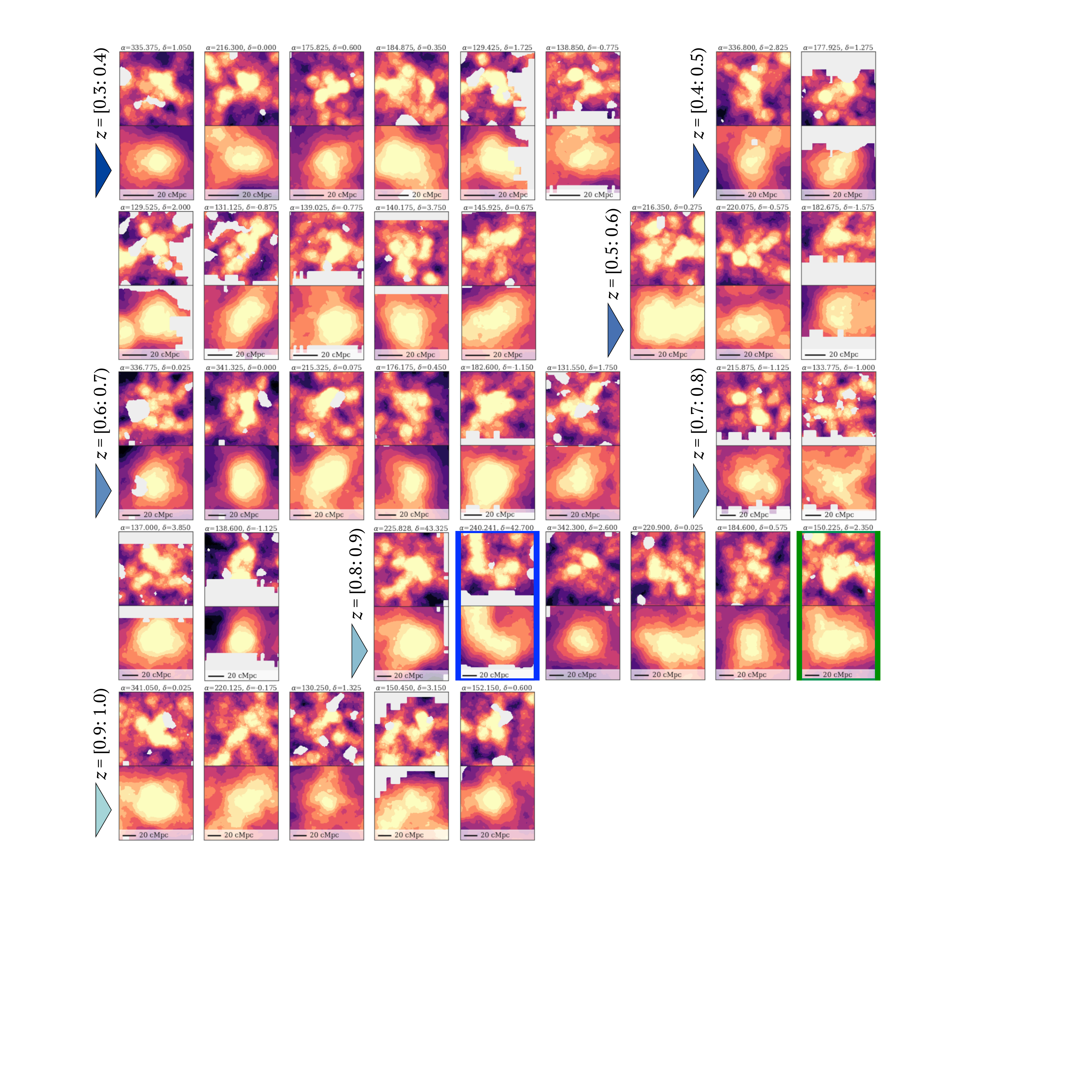}
    \caption{
    Examples of colossal over-densities at each redshift slice showing the number 
    excess of $>3\sigma$ (roughly corresponding to the density variance of 
    $\delta\gtrsim0.4$) at the density peak. For each panel of $2\times2$ sq. 
    deg, upper and lower maps show the density map within $r=10$ and 30 arcmin 
    apertures as in Fig.~\ref{fig6} and \ref{fig7}, respectively.
    The central coordinates are described on top of each panel. 
    At $z=[0.8,0.9)$, blue and green highlights are associated with the CL1604 and 
    the COSMOS superclusters \citep{Gal2008,Paulino-Afonso2018}, respectively.
    The black horizontal bar shown at the bottom of each panel indicates the 
    projected distance of 20 cMpc.
    }
    \label{fig9}
\end{figure*}

Furthermore, one of those lying at $z=$ 0.8 -- 0.9 in W03 GAMA09H 
($\alpha=150^\circ$, $\delta=2^\circ$; Fig.~\ref{fig9}) is likely associated with 
a COSMOS supercluster at $z=$ 0.8 -- 0.9 \citep{Paulino-Afonso2018}. We find that 
about a half of the COSMOS field \citep{Scoville2007} is covered by $>2$ sigma 
over-densities in $n_{r=30^\prime}$ at $z=$ 0.8 -- 0.9. In particular, the western 
side of the COSMOS supercluster shows the clustering $>4$ sigma density peaks in 
our photo-$z$ based analysis. 
This field is out of the survey area by \citet{Paulino-Afonso2018} unfortunately, 
and thus this large-scale over-dense region is not confirmed yet. 
A forthcoming deep galaxy evolution survey with Prime Focus Spectrograph (PFS) 
\citep{Takada2014} will explore the extended COSMOS region, including this 
interesting field in the near future. 

We detect $\sim50$ over-densities comparable to such supercluster-embedded regions 
given seven redshift slices while some of them spatially overlap with each other 
on the sky. Figure~\ref{fig9} summarises representative examples showing greater than three 
sigma excesses at the density peaks in the projected density distributions. 
Spectroscopic follow-up observations are required to confirm these enormous 
over-densities, though we evaluate their potential mass contents based on the 
cosmological simulation and the weak lensing analysis in the discussion 
(\S\ref{s4}). The density catalogue available online provides the number densities 
broken down into $\Delta z=0.02$, allowing more flexible supercluster search (see 
Appendix~\ref{a1}). 

Besides, the continuous wide-field coverage over $\gtrsim40$ sq. degrees in each 
field allows us to find widely-spread under-dense structures like cosmic troughs 
out to $z=1$ as well as over-densities. 
A cosmic trough is defined as under-dense circles of a fixed radius over a wide 
redshift slice \citep{Gruen2016}, which is one of promising probes of cosmic 
voids for the wide-field data. 
However, the currently available data may remain insufficient to probe supervoids 
with a scale of $\gtrsim400$ cMpc \citep{Higuchi2018} as discussed in 
relation with the cold spot of the cosmic microwave background 
\citep{Szapudi2015,Finelli2016}. High-quality imaging data from the Subaru 8.2m 
Telescope will provide an important insight into the in-situ dominated galaxy 
evolution. While studying the environmental dependence of galaxy properties over 
the survey fields is out of the scope of this paper, the forthcoming paper II 
will delve into individual galaxies residing in such unique environments. 

\begin{figure*}
    \centering
	\includegraphics[width=.95\textwidth]{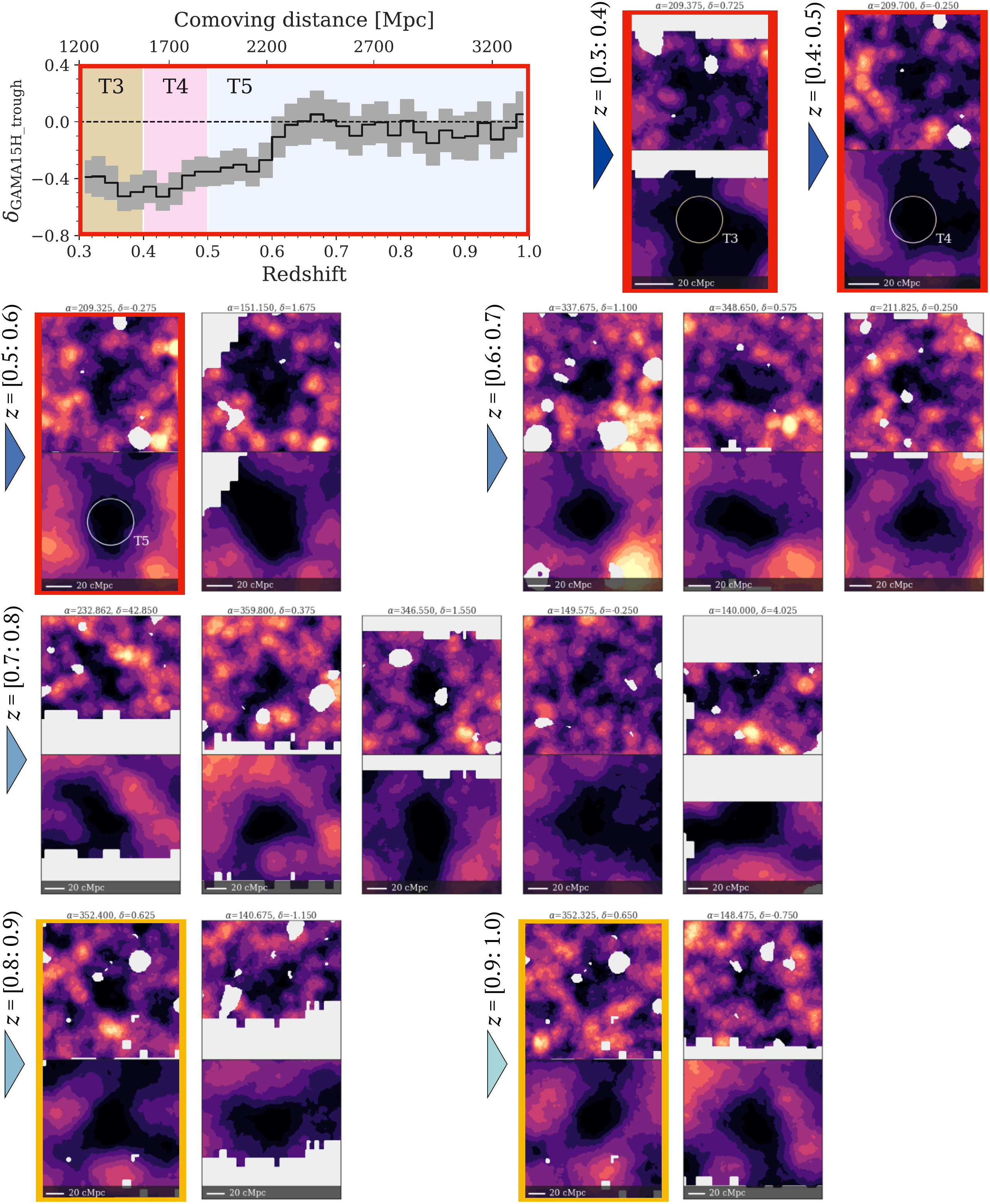}
    \caption{
    Examples of colossal under-densities at each redshift slice showing 
    $>2.5\sigma$ deficit ($\delta\lesssim-0.3$, $\delta$ is the density variance) 
    at the density trough. Each upper and lower maps are the cutouts of $3\times3$ 
    sq. deg from the density map within $r=10$ and 30 arcmin apertures as in 
    Fig.~\ref{fig6} and \ref{fig7}, respectively. 
    The central coordinates are denoted on top of each panel. 
    The three under-dense fields at $z=$ [0.3:0.4) framed by red colours are 
    located within the same region on the sky. The upper left figure indicates the 
    density variance at those troughs (T3, T4, T5) as a function of redshift or 
    comoving radial distance. The grey-filled regions show 1 sigma deviations at 
    each redshift bin of $\Delta z=0.02$. 
    The yellow-frame sources at $z=$ 0.8 -- 1 in the bottom are covered by the 
    Deep layer of HSC-SSP \citep{Aihara2018}. 
    The white horizontal bar shown at the bottom of each panel indicates the 
    projected distance of 20 cMpc.
    }
    \label{fig10}
\end{figure*}

As a result, we detect a few significant under-densities at each redshift bin, 
some of which are highlighted in Figure~\ref{fig10}. One should note that in 
Figure~\ref{fig10} we only choose under-dense regions whose density troughs are 
less affected by the bright-star masks (effective area of $>70$ per cent). 
The most prominent diffuse structure at $z<0.5$ is found in the W04 GAMA15H field 
($\alpha=209.375$, $\delta=0.725$; see in Fig.~\ref{fig10}). The density trough 
shows 2.5 -- 2.7 sigma deficit in the density measurement within $r=30$ arcmin 
apertures, which corresponds to $\delta\sim-0.5$ in the variance of number 
densities at this redshift range. Here $\delta$ is defined by 
$(n_\mathrm{r}-n_\mathrm{r,mean})/n_\mathrm{r,mean}$ where $n_\mathrm{r}$ is a 
number density within a radius of 10 arcmin or 30 arcmin, and $n_\mathrm{r,mean}$ 
is the mean value of $n_\mathrm{r}$. 
Intriguingly, the significant under-dense 
structure persists up to the redshift bin of $z=$ [0.5:0.6), suggesting that the 
diffuse structure could be of an extent $\sim$1,000 cMpc along the line of sight 
(see the top left panel in Fig.~\ref{fig10}). 
Another notable under-dense regions is seen in the W05 VVDS field at $z=$ 0.8 -- 
1 (Fig.~\ref{fig10}), which is in part covered by the Deep layer of HSC-SSP. 
Coming multi-band photometry at 0.3 -- 5 $\mu$m in the Deep layer 
\citep{Aihara2018,Sawicki2019,Moutard2020,Steinhardt2014} will enable detailed 
research in physical properties of galaxies living in such a vastly empty regions. 

Since the discovered enormous under-densities are widely spread over the scale of 
degrees, next-generation wide-field spectrographs are crucially important to 
obtain better constraints on their properties. The PFS on the Subaru Telescope 
\citep{Takada2014} will be able to achieve detailed three-dimensional mapping on 
cosmological scales in practical observing time, by its extremely-wide area 
coverage with a field-of-view of 1.38 degree diameter and simultaneous wavelength 
range at 0.38 -- 1.26 $\mu$m. 
Another solution to confirm these structures may be the convergence maps from 
weak lensing analyses as recently reported by \citet{Davies2020}, though it 
remains quite challenging to overcome galaxy shape noise on the several $\sim$ 
tens arcmin scale with the HSC-SSP data. 
Furthermore, \citet{Higuchi2018} have reported that a weak lensing analysis based 
on the on-going program like HSC-SSP can probe a supervoid at $z\sim0.2$ with a 
radius of $\sim300$ cMpc and a density contrast $\delta\sim-0.3$ at the trough, if 
such supervoids are present in the survey field. However, the currently available weak 
lensing data are largely restricted to the survey area of HSC-SSP PDR1, which is 
more patchy and has much smaller field coverage of 136.9 sq. deg compared to PDR2 
and the survey footprints ($\sim$1,400 sq. deg). We thus leave such weak lensing 
approaches to future work. Instead, this paper carries out the stacking weak 
lensing analysis, as described in the discussion section (\S\ref{s43}). 

\section{Verifying the colossal structures}
\label{s4}

It is quite important to assess the typical total masses and constrain variance of 
dark matter densities of the discovered enormous over- and under-dense structures. 
Such a quantification also works for more practical use of our density map 
catalogue (Appendix~\ref{a1}). In this context, we apply two approaches for 
the significant over- and under-density regions detected by this work: evaluating 
massive dark matter haloes associated with the over- and under-densities based on 
a mock galaxy catalogue from a cosmological simulation (\S\ref{s41}) and the 
weak lensing shear measurement (\S\ref{s43}) to address their underlying masses 
and densities as in the following sections.


\subsection{Mass distribution inferred from the simulation}
\label{s41}

At first, we specifically consider distributions of massive dark matter haloes 
across different environments inferred from the projected density map by using 
the mock galaxy catalogue ($i<23$ mag) from a cosmological simulation. We adopt 
the data-set from the all-sky gravitational lensing simulation created by 
\citet{Takahashi2017,Shirasaki2017}. 
Their simulation data excellently match the scientific motivations of this paper in 
the following respects: (i) the data cover a sufficiently large cosmological 
volume, and the area coverage can demonstrate the HSC-SSP Wide layer thanks to the all-sky ray-tracing capability; (ii) the 
data based on the weak lensing simulation for the HSC-SSP allow us to conduct a 
cross-comparison of the target regions between the observation and the simulation 
from the aspects both of galaxy distributions and weak lensing signals (see 
\S\ref{s43}). 

The details of the full-sky ray-tracing simulation are examined by 
\citet{Takahashi2017}. But we here summarise only the basic parts of the 
simulation relevant to our analysis. The simulation is based on the $N$-body code, 
{\tt GADGET2} \citep{Springel2005a} with 14 different boxes with size lengths of 
450 -- 6300 $h^{-1}$ Mpc in steps of 450 $h^{-1}$ Mpc to create lensing map at 
different source redshifts (\citealt[table 1]{Takahashi2017}). In each simulation 
box of $2048^3$ particles, they perform six independent realisations with the 
initial linear power spectrum based on the Code for Anisotropies in the Microwave 
Background \citep[{\tt CAMB}]{Lewis2000}. The light-ray path and the magnification 
on the lens planes are derived  for 38 different source redshifts from 0 to 5.3 
by using the multiple-plane gravitational lensing algorithm 
{\tt GRay-Trix}\footnote{\url{http://th.nao.ac.jp/MEMBER/hamanatk/GRayTrix/}} 
\citep{Hamana2001,Shirasaki2015}. The particles are positioned on to lens shells 
with a width of 150 $h^{-1}$ Mpc in the {\tt HEALPix} coordinates 
\citep{Gorski2005}. 
In order to reasonably evaluate the covariances of observables, they select 18 
observer's positions within each simulation box and increase the number of 
realisations. We refer readers to \citet[appendix C]{Shirasaki2015} for more 
details of the ray-tracing model. 
The simulation data include a dark matter halo catalogue from the $N$-body 
simulation using {\tt ROCKSTAR} \citep{Behroozi2013a} where a halo is defined as 
a group of more than 50 gravitationally bound particles. The simulation well 
resolves massive dark matter haloes of virial masses $\gtrsim10^{13}$ M$_\odot$ 
up to $z\sim1.2$ \citep{Takahashi2017}. 

Based on such a large-volume cosmological simulation, we establish one realisation 
suite specifically designed to mimic the survey field used in this work ($\sim360$ 
sq. deg) to check halo distributions surrounding the widely spread over- and 
under-dense structures. 
We generate $i$-band magnitude limited sources ($i<23$) at $z=$ 0.3 -- 1 from the 
over-density map of the dark matter particles. To keep consistency with our HSC 
sample, we adopt the large-scale galaxy bias $b(z)$, which is derived by 
\citet[eq.~4.11 and 4.12]{Nicola2020} based on the HSC PDR1 data 
\citep{Aihara2018a}. Assuming the linear galaxy bias may not work on the small 
scale $<10$ cMpc, and thus, this section mainly focuses on the results based on 
the density estimation with $r=30$ arcmin apertures (corresponding to $r=10$ cMpc 
at $z=0.3$). 
Since mock galaxy sources distribute on the celestial sphere corresponding to our 
survey area, we can conduct the density measurement in the same manner as 
for the observational data (\S\ref{s31}). While we do not need to perform the mask 
correction for the mock data, the edges of the survey fields are removed for the 
sake of simplifying the calculation. 
One should note that we here do not consider photo-$z$ uncertainties. However, we 
stress that photo-$z$ errors, and foreground and background contaminants do not cause 
a systematic effect on the density measurement but produce some scatters (see a 
detailed description in Appendix~\ref{a2}).

\begin{figure*}
    \centering
	\includegraphics[width=.95\textwidth]{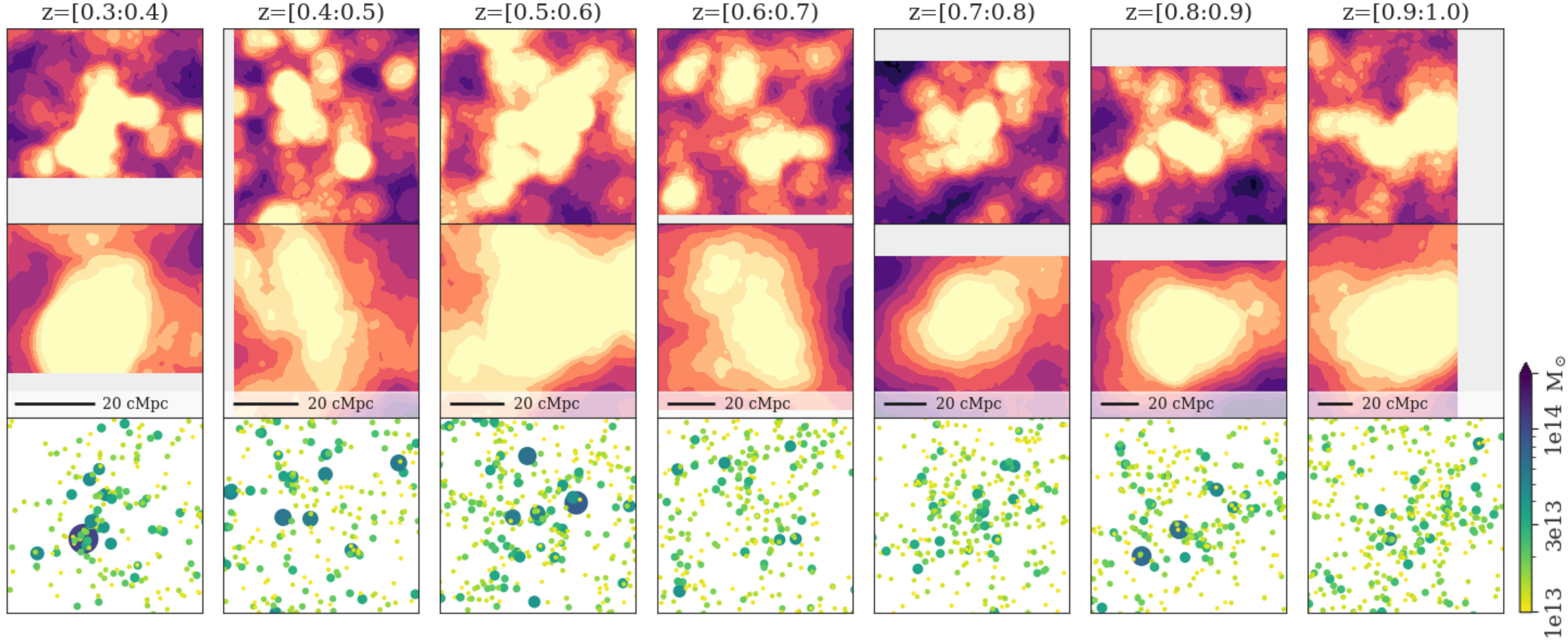}
    \caption{
    Examples of over-densities in each redshift slice of $\Delta z=0.1$ from 
    $z=$ [0.3:0.4) (the left panels) to $z=$ [0.9:1.0) (the right panels), which 
    are chosen from the projected density map based on the mock catalogue. 
    As in Figure~\ref{fig9}, each top and middle maps respectively show the 
    $2\times2$ deg$^2$ cutouts from the density map within $r=10$ and 30 arcmin 
    apertures. 
    Bottom panels show spatial distributions of massive dark matter haloes 
    (M$_\mathrm{vir}>1\times10^{13}$ M$_\odot$) within the redshift intervals 
    where colours and sizes of symbols are scaled depending on virial masses of 
    haloes. 
    }
    \label{fig11}
\end{figure*}

\begin{figure*}
    \centering
	\includegraphics[width=.95\textwidth]{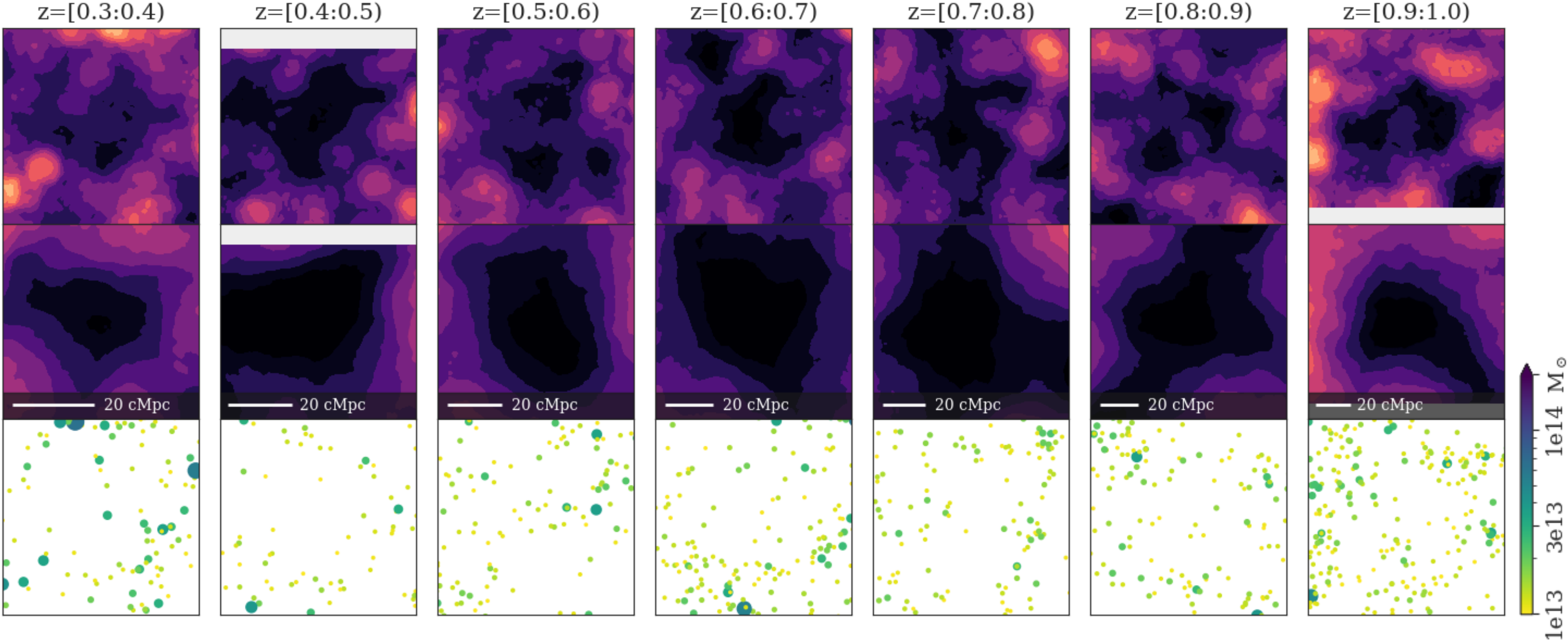}
    \caption{
    Same as Figure~\ref{fig11} but for under-dense regions.
    }
    \label{fig12}
\end{figure*}

\begin{figure*}
    \centering
	\includegraphics[width=.98\textwidth]{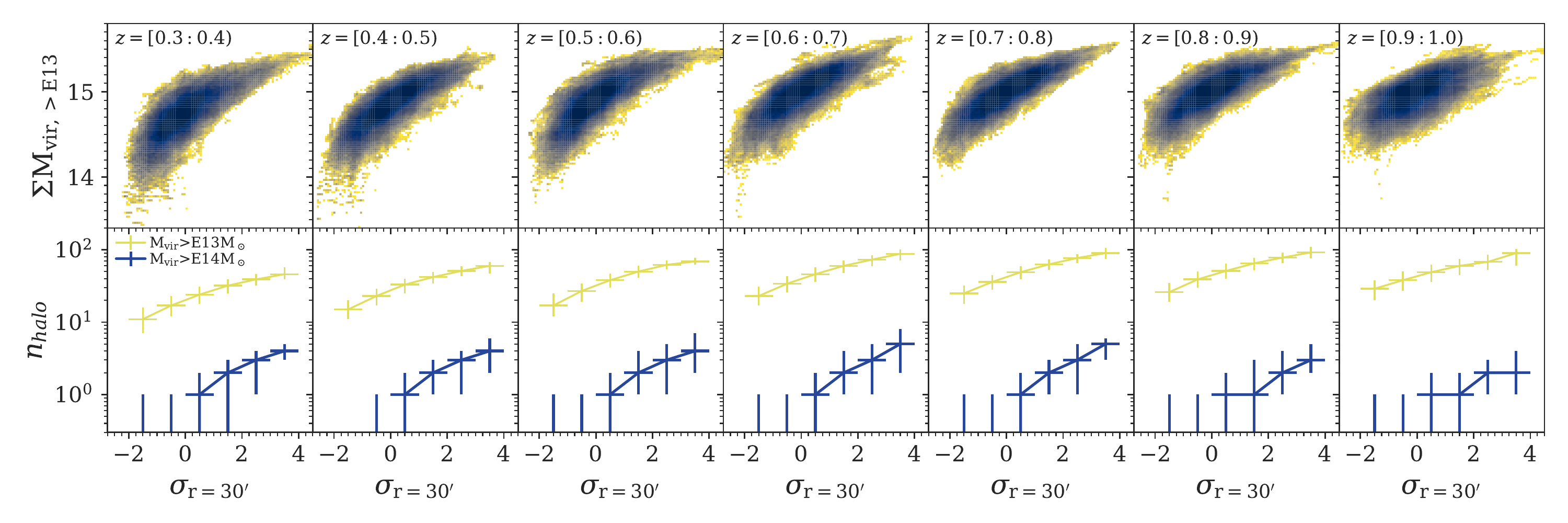}
    \caption{
    Upper panels indicate total masses of massive dark matter haloes 
    (M$_\mathrm{vir}>1\times10^{13}$ M$_\odot$) as a function of the density 
    excess ($\sigma_{r=30^\prime}$) within $r=30$ arcmin apertures at $z=0.3$ -- 
    1 in step with $\Delta z=0.1$ from left to right. 
    Lower panels show the number of massive haloes locating within the 
    corresponding apertures of $\Delta z=0.1$ and $r=30$ arcmin in each field. 
    Yellow and blue plots mean the number of massive haloes with 
    M$_\mathrm{vir}>1\times10^{13}$ M$_\odot$ and $>1\times10^{14}$ M$_\odot$, 
    respectively.
    Each point and errorbar indicate the median values and 68 percentiles in 
    step with $\Delta\sigma_{r=30^\prime}=1$ from $\sigma_{r=30^\prime}=-2$ to 
    $\sigma_{r=30^\prime}=4$. 
    }
    \label{fig13}
\end{figure*}

As in Figure~\ref{fig9} and \ref{fig10}, Figure~\ref{fig11} and ~\ref{fig12} 
respectively highlight some large-scale over- and under-dense regions detected in 
the `mock' projected density maps. The figures also show spatial 
distributions of massive dark matter haloes with 
M$_\mathrm{vir}>1\times10^{13}$ M$_\odot$ in each redshift space. 
These figures demonstrate that significant over-densities are associated with more 
massive dark matter haloes while the density troughs involve few massive haloes. 
More statistical tests about halo mass contents appear in Figure~\ref{fig13}, 
which shows the total mass of massive dark matter haloes $>10^{13}$ M$_\odot$ 
embedded within $r=30$ arcmin apertures as a function of the number density 
excess. Positive correlations between the number densities and the total 
halo masses suggest that our density measurement well traces the large-scale 
contrasts of halo distributions. Furthermore, the figure suggests that over-dense 
regions with $\sigma_{r=30^\prime}>3$ would host more than 1 -- 3 cluster-scale 
haloes with M$_\mathrm{vir}>1\times10^{14}$ M$_\odot$. This is consistent with a 
picture of superclusters as in the density peaks around the CL1604 and the 
COSMOS superclusters (\S\ref{s32}). On the other hand, the mock data suggest about 
70 per cent of the density troughs with $\sigma_{r=30^\prime}<-2$ do not involve 
massive haloes with M$_\mathrm{vir}>1\times10^{13}$ M$_\odot$ within the range of 
30 arcmin at each redshift. Thus, such significant under densities would be good 
candidates of cosmic voids with radii of $\gtrsim10$ cMpc \citep{Higuchi2013}.

\subsection{Basics of weak-lensing shear measurement}
\label{s42}

We then attempt to constrain the typical total masses of discovered over- and 
under-dense region by employing direct weak lensing measurements. Since it is 
not realistic to expect any significant detection for individual under-density 
regions, we adopt a stacking technique 
\citep{Amendola1999,Mandelbaum2006,Okabe2010,Oguri2012,Higuchi2013,Higuchi2016}. 
Before proceeding, we briefly overview the basics of the weak lensing analysis. 

An isotropic stretching describes gravitational lensing effects, i.e., convergence 
$\kappa$ and an anisotropic distortion called cosmic shear $\gamma_1$ and 
$\gamma_2$ with the magnification matrix $A$:
\begin{equation}
  A_{ij}
  = 
  \frac{\partial\beta^i}{\partial\theta^j}
  \equiv
  \left(
    \begin{array}{cc}
      1-\kappa-\gamma_1 & -\gamma_2 \\
      -\gamma_2 & 1-\kappa+\gamma_1
    \end{array}
  \right)
\end{equation}
where $\theta$ is the lensed angular sky position of a source object and 
$\beta$ is the un-lensed true position. 
Practically, the decomposition of the shear field, the tangential distortion  
$\gamma_+$ and the cross distortion $\gamma_\times$ components are the only 
requirements for this work. These are written by the following form, 
\begin{equation}
  \left(
    \begin{array}{c}
      \gamma_+ \\
      \gamma_\times
    \end{array}
  \right)
  = 
  \left(
    \begin{array}{cc}
      -\cos2\eta & -\sin2\eta \\
      -\sin2\eta & \cos2\eta
    \end{array}
  \right)
  \left(
    \begin{array}{c}
      \gamma_1 \\
      \gamma_2
    \end{array}
  \right)
\end{equation}
where $\eta$ is the angle between axis $\alpha$ and $\theta$ on the 
$\alpha$--$\beta$ plane. The tangential shear becomes positive (negative) when 
galaxy shape tangentially (radially) deformed with respect to the lens centre. 
Assuming an axisymmetric lensing profile, the tangential shear contains all 
the information from lensing, while the cross shear value is ideally zero and 
thus this can be used to test residual systematic effects. In our stacking 
analysis, we measure the mean tangential shear profile as a function of 
radial distances, $\langle\gamma_+\rangle$ at radius $\theta_i$ from a 
targeting area 
\citep{Bartelmann2001},
\begin{equation}
\langle\gamma_+\rangle(\theta_i) = \tilde{\kappa}(\theta<\theta_i)
- \langle\kappa\rangle(\theta_i),
\end{equation}
where $\tilde{\kappa}$ is the mean convergence within $\theta_i$. The convergence 
traces the Laplacian of the scalar potential, $\psi(\theta)$ as follows,
\begin{equation}
\begin{split}
\kappa(\theta) &= \frac{1}{2}\nabla^2\psi(\theta),\\
&= \frac{3H_0^2\Omega_{M}}{2c^2}\frac{D_{ls}D_l}{D_s a_l^3}\int\mathrm{d}z\delta(\theta,z),\\
\tilde{\kappa}(\theta<\theta_i) &= 
\frac{1}{\pi\theta^2}\int_{\theta\le\theta_i}\mathrm{d}\theta^\prime\kappa(\theta^\prime),
\end{split}
\end{equation}
where $\delta(\theta,z)$ is the local density fluctuation at the line-of-sight 
distance $z$ ($\rho(\theta,z)/\tilde{\rho}(z)-1$). $D_s$, $D_l$ and $D_ls$ are 
the angular diameter distances as described below,
\begin{equation}
D_s=a_s\chi_s, \\ D_l=a_l\chi_l, \\ D_{ls}=a_s(\chi_s-\chi_l).
\end{equation}
$\chi_s$ and $\chi_l$ are the comoving distance from an observer to source and 
lens planes, and $a_s$ is the scale factor at the source position. 

Therefore, measuring the mean tangential shear profile 
$\langle\gamma_+\rangle$ centring on the targets can provide direct 
insights into their typical density contrast $\delta$ without any knowledge of the 
galaxy bias. 
Also, such weak lensing signals give complementary information to what our 
realistic mocks provide in the previous subsection (\S\ref{s41}). On top of that, 
since our mock data from the ray-tracing simulation includes the galaxy shape 
catalogue for the weak lensing analysis \citep{Shirasaki2019}, we can cross-check 
results within all these data as discussed in the following section.


\subsection{Cross-check with a weak lensing analysis}
\label{s43}

The weak lensing stacking analysis provides direct insights into the density 
contrast centred on the target regions. We employ the first-year shear catalogue 
of HSC-SSP that is available in 
public.\footnote{\url{https://hsc-release.mtk.nao.ac.jp/doc/index.php/s16a-shape-catalog-pdr2/}}
The data cover 136.9 sq. degrees of the HSC-SSP Wide layer taken during March 2014 
and April 2016 \citep{Mandelbaum2018}. 
To test the validation of our weak lensing analysis, we apply the same procedure 
to the mock galaxy shape catalogue \citep{Shirasaki2019} which is based on the 
cosmological weak lensing simulation \citep{Takahashi2017} introduced in 
\S\ref{s41}. While the mock catalogue consists of in total 2,268 realisations of 
each HSC PDR1 region from 108 full-sky simulation runs with 21 rotations 
(i.e., $108\times21=$ 2,268), this work picks up 30 realisations out of 108 full-sky simulation catalogues. Since we are motivated to assess the typical 
density contrasts and variations around the large over- and under-densities but 
not aiming for cosmological constraints, 30 realisation data should be enough to 
check systematic trends and scatters of the lens shear signals for our purpose. 

The detailed methodology of the weak lensing stacking is given by 
\cite{Higuchi2013,Higuchi2019}; Higuchi et al. in preparation, while we describe 
some important notes in our approach below. 
In the shear measurement, we do not adopt any redshift cut for selecting 
background galaxies and use all of the galaxies in the shape catalogue. The errors 
in the shear measurement, such as patchy survey footprints of the HSC survey would 
generate systematics. To subtract such effects, we subtract lensing signals at 
random points from the shear profiles for the over- and under-dense regions 
\citep{Gruen2016}. The errors for the observational results are estimated 
from the 30 realisations of the mocks (Higuchi et al. in preparation). 

\begin{figure}
    \centering
	\includegraphics[width=.95\columnwidth]{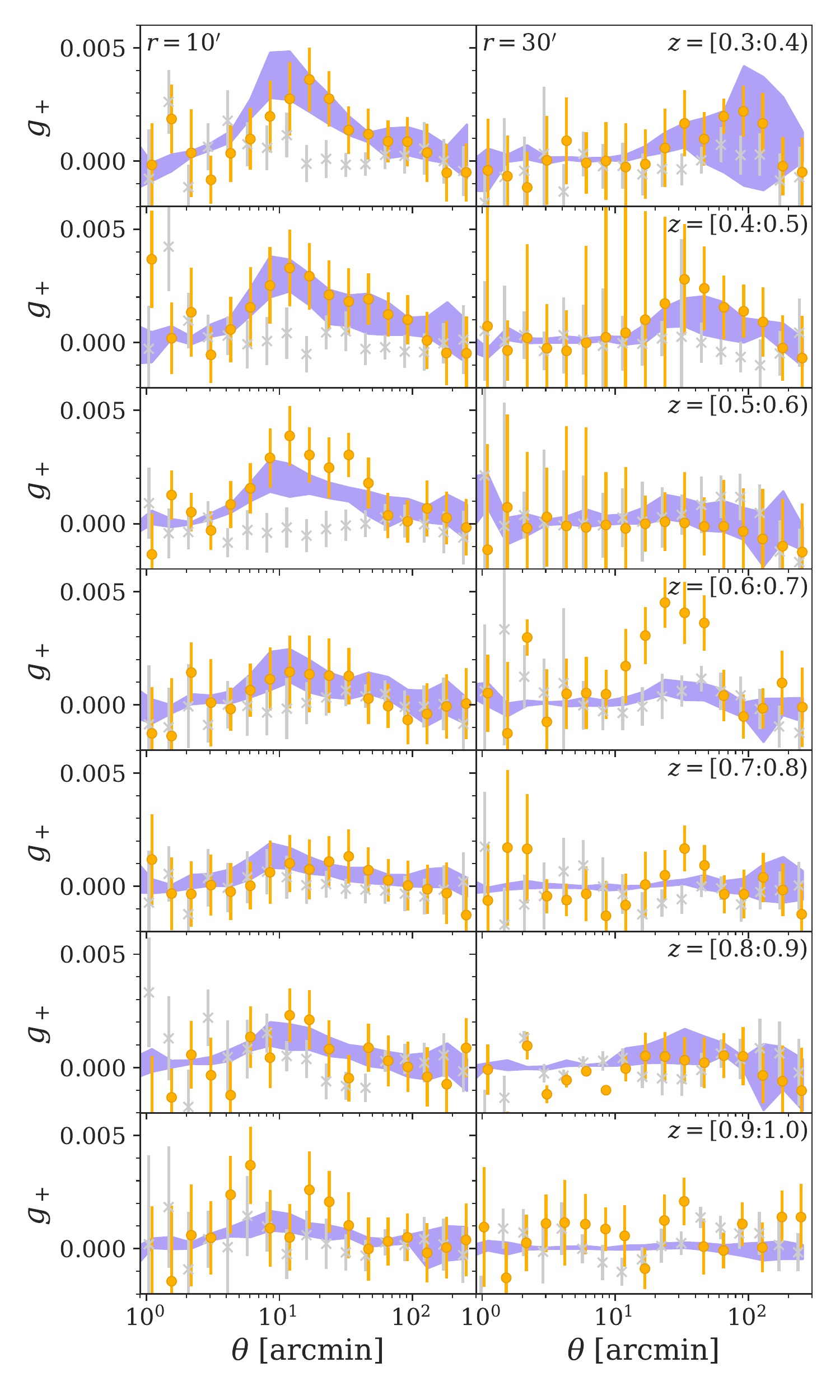}
    \caption{
    The orange circles and the grey cross symbols show the mean tangential and 
    cross shear profiles from the weak lensing stacking for the over-densities 
    with the number excess of $>3\sigma$ at $z=$ 0.3 -- 1 in step of 
    $\Delta z=0.1$ from top to bottom. Left and right panels respectively show
    the stacked lens signals for over-dense regions selected by $r=10$ and 30 
    arcmin apertures. 
    The purple region indicates 68th percentiles of the mean tangential profiles 
    from the mock shape catalogue based on 30 realisations.
    }
    \label{fig14}
\end{figure}

\begin{figure}
    \centering
	\includegraphics[width=.95\columnwidth]{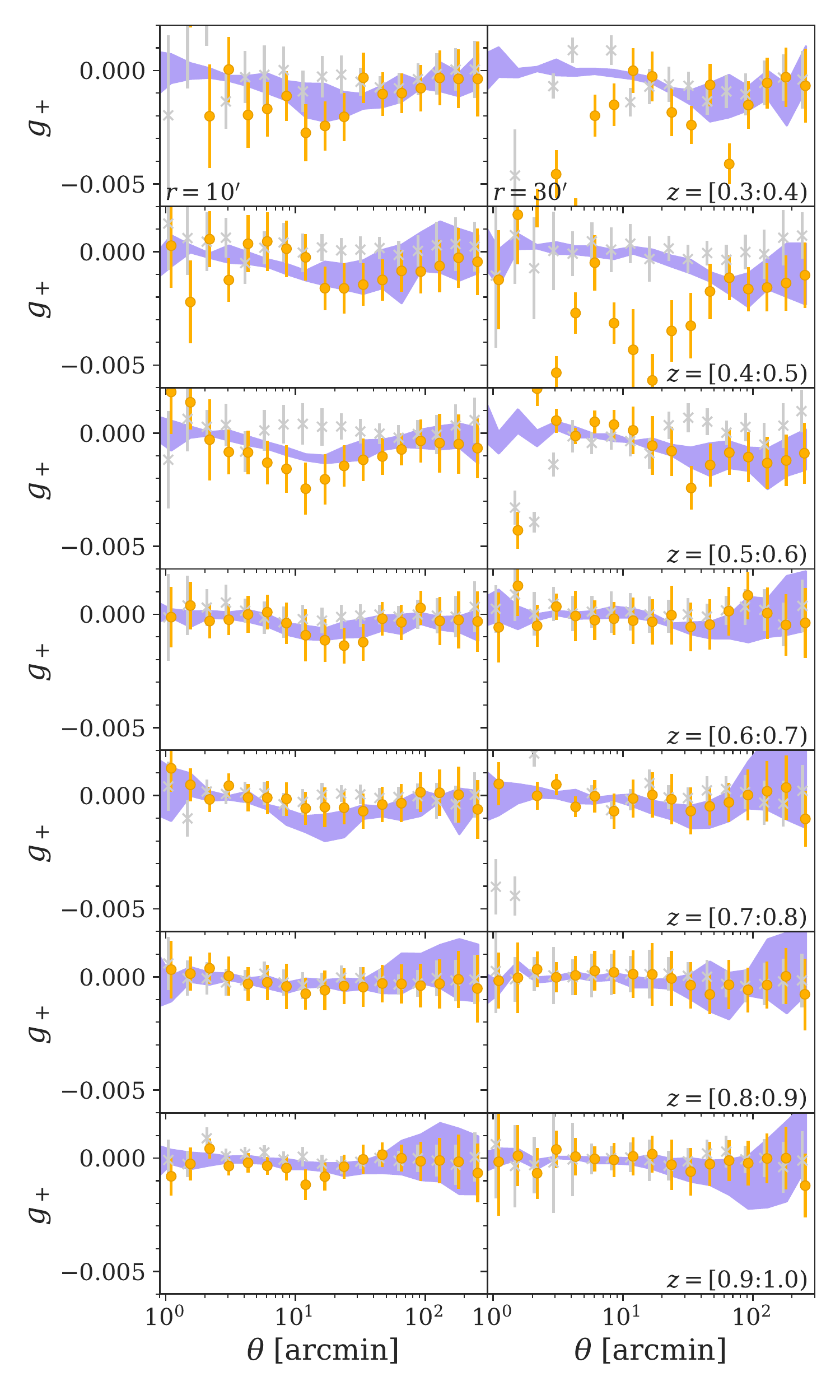}
    \caption{
    Same as Figure~\ref{fig14} but the mean tangential shear profiles for the 
    under-dense regions with the number deficit of $<-1.5\sigma$.
    }
    \label{fig15}
\end{figure}

The resultant mean tangential shear profiles for both observational and simulation 
data are presented in Figure~\ref{fig14} and \ref{fig15}. In the stacking 
analysis, we select the over- and under-dense regions at each redshift slice 
showing the number excesses of $>3\sigma$ and $<-1.5\sigma$, respectively.  
The measured mean tangential shear profiles of the observation-based density peaks 
and troughs are generally consistent with the results from the mocks within the 
margin of error. However, the signals are too noisy and/or small to obtain 
meaningful information at $z>0.6$. 
A more careful redshift selection for background galaxies would be needed to 
obtain a better signal-to-noise ratio of the lensing profiles at $z>0.6$, but we 
leave it for future work.
Shapes of the detected weak lens profiles also show an agreement with the 
observational and analytical results obtained in the Dark energy survey 
\citep{Gruen2016,Gruen2018}.

Ideally speaking, it is expected that over- and under-densities respectively show 
positive and negative shear signals at $\theta$ corresponding to the aperture 
radius. Indeed, in the $r=10$ arcmin aperture measurement, the mock results 
present the signal peaks and troughs around $\theta\sim10$ arcmin, while the 
signal strengths become weaker at higher redshift bins (Fig.~\ref{fig14} and 
\ref{fig15}). The signal detection and the consistency between the observation and 
the mock data at $z\lesssim0.6$ suggest that our density estimation well probe 
analytical over- and under-densities as generated by the cosmological simulation. 
On the other hand, as inferred from the mock samples, the larger aperture ($r=30$ 
arcmin) density estimation is less sensitive to the local density contrast than 
the $r=10$ arcmin aperture, weakening the shear signals which are hardly detected 
in the current data. Particularly, some outlier signals in the observations (e.g., 
$\theta\sim20$ arcmin at $z=[0.4:0.5)$ in Fig.~\ref{fig15}) may be due to the 
limited field coverage of the previous PDR1. 
The future release of the galaxy shape catalogue for the entire survey footprint 
($\sim$1,400 sq. deg) will be able to improve signal-to-noise ratios by a factor 
of $\sim3$, allowing better constraints on the weak lensing shear profiles.

\section{Conclusion}
\label{s5}

Based on the HSC-SSP PDR2 data, we estimate projected number densities of $i$-band 
selected sources ($i<23$ mag) from $z=0.3$ to $z=1$ in step with $\Delta z=0.1$, 
within two different aperture areas ($r=10$ and 30 arcmin). This paper mainly 
focuses on the large-scale density contrasts based on the latter aperture size, 
though the former one would be more practical for the purpose of spectroscopic 
follow-up observations. 
The wide-field and deep HSC data successfully demonstrate the large-scale 
structure over 360 sq. degree fields, including the confirmation of known 
superclusters at $z=$ 0.8 -- 1. By applying the same technique to the mock galaxy 
catalogue generated from the all-sky ray-tracing cosmological simulation, we find 
that the projected over-densities well trace the total masses of embedded massive 
dark matter haloes at each redshift slice. While the significant density peaks are 
expected to host one or more cluster-scale haloes at corresponding redshifts, the 
density troughs, top candidates of distant cosmic voids, would remarkably lack the 
matters and massive dark matter haloes. The significant density peaks and troughs 
at $z\lesssim0.6$ are also confirmed with the mean lens shear signals from the 
weak lensing stacking analysis. 

Thus far, intensive spectroscopic follow-up observations have been executed to 
a few superclusters at $z\sim1$ 
\citep{Gal2008,Lubin2009,Lietzen2016,Paulino-Afonso2018}, though an extensive and 
comprehensive spectroscopic database has not yet been formed for mid and high 
redshift superclusters and voids. 
Thus a massive spectroscopic survey, like DESI \citep{DESICollaboration2016}, PFS 
\citep{Takada2014}, MOONS \citep{Cirasuolo2011} and 4MOST \citep{DeJong2019} is 
imperative to address detailed structures of galaxies associated with the colossal 
over- and under-densities. 
Besides, eROSITA \citep{Merloni2012}, the Vera C. Rubin Observatory 
\citep{Ivezic2019}, Euclid \citep{Amendola2018}, and the Roman Space Telescope 
\citep{Dore2019} are crucial to accumulate systematic information of such enormous 
structures or extend samples to the earlier universe. 

Lastly, we should note that there remains substantial scope for improvement in our 
analyses, e.g., mask correction and removal of contaminants that are not perfect 
in PDR2 \citep{Aihara2019}. 
These caveats will be addressed with the third or later data release, as our data 
reduction pipeline keeps improving the data quality.


\section*{Acknowledgements}

We thank the anonymous reviewer for the helpful comments.
Analyses are in part conducted with the assistance of the Tool for OPerations on 
Catalogues And Tables ({\tt TOPCAT}; \citealt{Taylor2005}), {\tt Astropy}, a 
community-developed core Python package for Astronomy \citep{Robitaille2013}, and 
{\tt pandas}, Python Data Analysis Library \citep{McKinney2010}. 
R.S. acknowledges the support from Grants-in-Aid for Scientific Research (KAKENHI; 
19K14766)
through the Japan Society for the Promotion of Science (JSPS).
This work was in part supported by Grant-in-Aid for Scientific Research on 
Innovative Areas from the MEXT KAKENHI Grant Number (18H04358, 19K14767).

This paper is based on data collected at the Subaru Telescope and retrieved from 
the HSC data archive system, which is operated by Subaru Telescope and Astronomy 
Data Center at NAOJ. Data analysis was in part carried out with the cooperation of 
Center for Computational Astrophysics, NAOJ.
The HSC collaboration includes the astronomical communities of Japan and Taiwan, 
and Princeton University. The HSC instrumentation and software were developed by 
NAOJ, the Kavli Institute for the Physics and Mathematics of the Universe (Kavli 
IPMU), the University of Tokyo, the High Energy Accelerator Research Organization 
(KEK), the Academia Sinica Institute for Astronomy and Astrophysics in Taiwan 
(ASIAA), and Princeton University. Funding was contributed by the FIRST program 
from the Japanese Cabinet Office, the Ministry of Education, Culture, Sports, 
Science and Technology, the Japan Society for the Promotion of Science, Japan 
Science and Technology Agency, the Toray Science Foundation, NAOJ, Kavli IPMU, 
KEK, ASIAA, and Princeton University. This paper makes use of software developed 
for the Large Synoptic Survey Telescope (LSST). We thank the LSST Project for 
making their code available as free software at \url{http://dm.lsst.org}.

The Pan-STARRS1 Surveys (PS1) and the PS1 public science archive have been made 
possible through contributions by the Institute for Astronomy, the University of 
Hawaii, the Pan-STARRS Project Office, the Max Planck Society and its 
participating institutes, the Max Planck Institute for Astronomy, Heidelberg, and 
the Max Planck Institute for Extraterrestrial Physics, Garching, The Johns 
Hopkins University, Durham University, the University of Edinburgh, the Queen's 
University Belfast, the Harvard-Smithsonian Center for Astrophysics, the Las 
Cumbres Observatory Global Telescope Network Incorporated, the National Central 
University of Taiwan, the Space Telescope Science Institute, the National 
Aeronautics and Space Administration under grant No. NNX08AR22G issued through 
the Planetary Science Division of the NASA Science Mission Directorate, the 
National Science Foundation grant No. AST-1238877, the University of Maryland, 
Eotvos Lorand University, the Los Alamos National Laboratory, and the Gordon and 
Betty Moore Foundation.
Y.H is supported by ALMA collaborative science research project 2018-07A.

The authors wish to recognise and acknowledge the very 
significant cultural role and reverence that the summit of Maunakea has always 
had within the indigenous Hawaiian community. We are most fortunate to have the 
opportunity to conduct observations from this mountain.


\section*{Data Availability}

The data underlying this article are all available under the public data release site of 
Hyper Suprime-Cam Subaru Strategic Program (\url{https://hsc.mtk.nao.ac.jp/ssp/data-release/}).




\bibliographystyle{mnras}
\bibliography{RS20a} 

\begin{thebibliography}{}
\makeatletter
\relax
\def\mn@urlcharsother{\let\do\@makeother \do\$\do\&\do\#\do\^\do\_\do\%\do\~}
\def\mn@doi{\begingroup\mn@urlcharsother \@ifnextchar [ {\mn@doi@}
  {\mn@doi@[]}}
\def\mn@doi@[#1]#2{\def\@tempa{#1}\ifx\@tempa\@empty \href
  {http://dx.doi.org/#2} {doi:#2}\else \href {http://dx.doi.org/#2} {#1}\fi
  \endgroup}
\def\mn@eprint#1#2{\mn@eprint@#1:#2::\@nil}
\def\mn@eprint@arXiv#1{\href {http://arxiv.org/abs/#1} {{\tt arXiv:#1}}}
\def\mn@eprint@dblp#1{\href {http://dblp.uni-trier.de/rec/bibtex/#1.xml}
  {dblp:#1}}
\def\mn@eprint@#1:#2:#3:#4\@nil{\def\@tempa {#1}\def\@tempb {#2}\def\@tempc
  {#3}\ifx \@tempc \@empty \let \@tempc \@tempb \let \@tempb \@tempa \fi \ifx
  \@tempb \@empty \def\@tempb {arXiv}\fi \@ifundefined
  {mn@eprint@\@tempb}{\@tempb:\@tempc}{\expandafter \expandafter \csname
  mn@eprint@\@tempb\endcsname \expandafter{\@tempc}}}

\bibitem[\protect\citeauthoryear{Abbott et~al.,}{Abbott
  et~al.}{2018}]{Abbott2018}
Abbott T. M.~C.,  et~al., 2018, \mn@doi [ApJS] {10.3847/1538-4365/aae9f0}, 239,
  18

\bibitem[\protect\citeauthoryear{Agarwal \& Feldman}{Agarwal \&
  Feldman}{2011}]{Agarwal2011}
Agarwal S.,  Feldman H.~A.,  2011, \mn@doi [MNRAS]
  {10.1111/j.1365-2966.2010.17546.x}, 410, 1647

\bibitem[\protect\citeauthoryear{Aihara et~al.,}{Aihara
  et~al.}{2018a}]{Aihara2018}
Aihara H.,  et~al., 2018a, \mn@doi [PASJ] {10.1093/pasj/psx066}, 70

\bibitem[\protect\citeauthoryear{Aihara et~al.,}{Aihara
  et~al.}{2018b}]{Aihara2018a}
Aihara H.,  et~al., 2018b, \mn@doi [PASJ] {10.1093/pasj/psx081}, 70, 1

\bibitem[\protect\citeauthoryear{Aihara et~al.,}{Aihara
  et~al.}{2019}]{Aihara2019}
Aihara H.,  et~al., 2019, \mn@doi [PASJ] {10.1093/pasj/psz103}, 71, 2

\bibitem[\protect\citeauthoryear{Alpaslan et~al.,}{Alpaslan
  et~al.}{2015}]{Alpaslan2015}
Alpaslan M.,  et~al., 2015, \mn@doi [MNRAS] {10.1093/mnras/stv1176}, 451, 3249

\bibitem[\protect\citeauthoryear{Amendola, Frieman  \& Waga}{Amendola
  et~al.}{1999}]{Amendola1999}
Amendola L.,  Frieman J.~A.,   Waga I.,  1999, \mn@doi [MNRAS]
  {10.1046/j.1365-8711.1999.02841.x}, 309, 465

\bibitem[\protect\citeauthoryear{Amendola et~al.,}{Amendola
  et~al.}{2018}]{Amendola2018}
Amendola L.,  et~al., 2018, \mn@doi [Living Rev. Relativ.]
  {10.1007/s41114-017-0010-3}, 21, 2

\bibitem[\protect\citeauthoryear{Baldry, Balogh, Bower, Glazebrook, Nichol,
  Bamford  \& Budavari}{Baldry et~al.}{2006}]{Baldry2006}
Baldry I.~K.,  Balogh M.~L.,  Bower R.~G.,  Glazebrook K.,  Nichol R.~C.,
  Bamford S.~P.,   Budavari T.,  2006, \mn@doi [MNRAS]
  {10.1111/j.1365-2966.2006.11081.x}, 373, 469

\bibitem[\protect\citeauthoryear{Bartelmann \& Schneider}{Bartelmann \&
  Schneider}{2001}]{Bartelmann2001}
Bartelmann M.,  Schneider P.,  2001, \mn@doi [Phys. Rep.]
  {10.1016/S0370-1573(00)00082-X}, 340, 291

\bibitem[\protect\citeauthoryear{Behroozi, Wechsler  \& Wu}{Behroozi
  et~al.}{2013}]{Behroozi2013a}
Behroozi P.~S.,  Wechsler R.~H.,   Wu H.-Y.,  2013, \mn@doi [ApJ]
  {10.1088/0004-637X/762/2/109}, 762, 109

\bibitem[\protect\citeauthoryear{Blanton et~al.,}{Blanton
  et~al.}{2017}]{Blanton2017}
Blanton M.~R.,  et~al., 2017, \mn@doi [AJ] {10.3847/1538-3881/aa7567}, 154, 28

\bibitem[\protect\citeauthoryear{Bosch et~al.,}{Bosch et~al.}{2018}]{Bosch2018}
Bosch J.,  et~al., 2018, \mn@doi [PASJ] {10.1093/pasj/psx080}, 70, 1

\bibitem[\protect\citeauthoryear{Cai, Padilla  \& Li}{Cai
  et~al.}{2015}]{Cai2015}
Cai Y.-C.,  Padilla N.,   Li B.,  2015, \mn@doi [MNRAS] {10.1093/mnras/stv777},
  451, 1036

\bibitem[\protect\citeauthoryear{{Cirasuolo}, {Afonso}, {Bender}, {Bonifacio},
  {Evans}, {Kaper}, {Oliva}  \& {Vanzi}}{{Cirasuolo}
  et~al.}{2011}]{Cirasuolo2011}
{Cirasuolo} M.,  {Afonso} J.,  {Bender} R.,  {Bonifacio} P.,  {Evans} C.,
  {Kaper} L.,  {Oliva} E.,   {Vanzi} L.,  2011, The Messenger, \href
  {https://ui.adsabs.harvard.edu/abs/2011Msngr.145...11C} {145, 11}

\bibitem[\protect\citeauthoryear{Clampitt, Cai  \& Li}{Clampitt
  et~al.}{2013}]{Clampitt2013}
Clampitt J.,  Cai Y.-C.,   Li B.,  2013, \mn@doi [MNRAS]
  {10.1093/mnras/stt219}, 431, 749

\bibitem[\protect\citeauthoryear{Colless et~al.,}{Colless
  et~al.}{2001}]{Colless2001}
Colless M.,  et~al., 2001, \mn@doi [MNRAS] {10.1046/j.1365-8711.2001.04902.x},
  328, 1039

\bibitem[\protect\citeauthoryear{Contarini, Ronconi, Marulli, Moscardini,
  Veropalumbo  \& Baldi}{Contarini et~al.}{2019}]{Contarini2019}
Contarini S.,  Ronconi T.,  Marulli F.,  Moscardini L.,  Veropalumbo A.,
  Baldi M.,  2019, \mn@doi [MNRAS] {10.1093/mnras/stz1989}, 488, 3526

\bibitem[\protect\citeauthoryear{Coupon, Czakon, Bosch, Komiyama, Medezinski,
  Miyazaki  \& Oguri}{Coupon et~al.}{2018}]{Coupon2018}
Coupon J.,  Czakon N.,  Bosch J.,  Komiyama Y.,  Medezinski E.,  Miyazaki S.,
  Oguri M.,  2018, \mn@doi [PASJ] {10.1093/pasj/psx047}, 70, 1

\bibitem[\protect\citeauthoryear{{DESI Collaboration} et~al.,}{{DESI
  Collaboration} et~al.}{2016}]{DESICollaboration2016}
{DESI Collaboration} et~al., 2016, arXiv e-prints, arXiv:1611.00036

\bibitem[\protect\citeauthoryear{{Davies}, {Paillas}, {Cautun}  \&
  {Li}}{{Davies} et~al.}{2020}]{Davies2020}
{Davies} C.~T.,  {Paillas} E.,  {Cautun} M.,   {Li} B.,  2020, arXiv e-prints,
  \href {https://ui.adsabs.harvard.edu/abs/2020arXiv200411387D}
  {arXiv:2004.11387}

\bibitem[\protect\citeauthoryear{Dekel \& Rees}{Dekel \&
  Rees}{1994}]{Dekel1994}
Dekel A.,  Rees M.~J.,  1994, \mn@doi [ApJ] {10.1086/187197}, 422, L1

\bibitem[\protect\citeauthoryear{Dor{\'{e}} et~al.,}{Dor{\'{e}}
  et~al.}{2019}]{Dore2019}
Dor{\'{e}} O.,  et~al., 2019, arXiv e-prints, arXiv:1904.01174

\bibitem[\protect\citeauthoryear{Douglass \& Vogeley}{Douglass \&
  Vogeley}{2017}]{Douglass2017}
Douglass K.~A.,  Vogeley M.~S.,  2017, \mn@doi [ApJ]
  {10.3847/1538-4357/834/2/186}, 834, 186

\bibitem[\protect\citeauthoryear{Driver et~al.,}{Driver
  et~al.}{2009}]{Driver2009}
Driver S.~P.,  et~al., 2009, \mn@doi [Astron. Geophys.]
  {10.1111/j.1468-4004.2009.50512.x}, 50, 5.12

\bibitem[\protect\citeauthoryear{Driver et~al.,}{Driver
  et~al.}{2011}]{Driver2011}
Driver S.~P.,  et~al., 2011, \mn@doi [MNRAS]
  {10.1111/j.1365-2966.2010.18188.x}, 413, 971

\bibitem[\protect\citeauthoryear{Eisenstein et~al.,}{Eisenstein
  et~al.}{2011}]{Eisenstein2011}
Eisenstein D.~J.,  et~al., 2011, \mn@doi [AJ] {10.1088/0004-6256/142/3/72},
  142, 72

\bibitem[\protect\citeauthoryear{Finelli, Garc{\'{i}}a-Bellido, Kov{\'{a}}cs,
  Paci  \& Szapudi}{Finelli et~al.}{2016}]{Finelli2016}
Finelli F.,  Garc{\'{i}}a-Bellido J.,  Kov{\'{a}}cs A.,  Paci F.,   Szapudi I.,
   2016, \mn@doi [MNRAS] {10.1093/mnras/stv2388}, 455, 1246

\bibitem[\protect\citeauthoryear{Furusawa et~al.,}{Furusawa
  et~al.}{2018}]{Furusawa2018}
Furusawa H.,  et~al., 2018, \mn@doi [PASJ] {10.1093/pasj/psx079}, 70, 1

\bibitem[\protect\citeauthoryear{Gal, Lemaux, Lubin, Kocevski  \& Squires}{Gal
  et~al.}{2008}]{Gal2008}
Gal R.~R.,  Lemaux B.~C.,  Lubin L.~M.,  Kocevski D.,   Squires G.~K.,  2008,
  \mn@doi [ApJ] {10.1086/590416}, 684, 933

\bibitem[\protect\citeauthoryear{Geller \& Hwang}{Geller \&
  Hwang}{2015}]{Geller2015}
Geller M.~J.,  Hwang H.~S.,  2015, \mn@doi [Astron. Nachrichten]
  {10.1002/asna.201512182}, 336, 428

\bibitem[\protect\citeauthoryear{Geller, Diaferio  \& Kurtz}{Geller
  et~al.}{2011}]{Geller2011}
Geller M.~J.,  Diaferio A.,   Kurtz M.~J.,  2011, \mn@doi [AJ]
  {10.1088/0004-6256/142/4/133}, 142, 133

\bibitem[\protect\citeauthoryear{Gorski, Hivon, Banday, Wandelt, Hansen,
  Reinecke  \& Bartelmann}{Gorski et~al.}{2005}]{Gorski2005}
Gorski K.~M.,  Hivon E.,  Banday A.~J.,  Wandelt B.~D.,  Hansen F.~K.,
  Reinecke M.,   Bartelmann M.,  2005, \mn@doi [ApJ] {10.1086/427976}, 622, 759

\bibitem[\protect\citeauthoryear{Gregory \& Thompson}{Gregory \&
  Thompson}{1978}]{Gregory1978}
Gregory S.~A.,  Thompson L.~A.,  1978, \mn@doi [ApJ] {10.1086/156198}, 222, 784

\bibitem[\protect\citeauthoryear{Gruen et~al.,}{Gruen et~al.}{2016}]{Gruen2016}
Gruen D.,  et~al., 2016, \mn@doi [MNRAS] {10.1093/mnras/stv2506}, 455, 3367

\bibitem[\protect\citeauthoryear{Gruen et~al.,}{Gruen et~al.}{2018}]{Gruen2018}
Gruen D.,  et~al., 2018, \mn@doi [Phys. Rev. D] {10.1103/PhysRevD.98.023507},
  98, 023507

\bibitem[\protect\citeauthoryear{Gunn, Hoessel  \& Oke}{Gunn
  et~al.}{1986}]{Gunn1986}
Gunn J.~E.,  Hoessel J.~G.,   Oke J.~B.,  1986, \mn@doi [ApJ] {10.1086/164317},
  306, 30

\bibitem[\protect\citeauthoryear{Hamana \& Mellier}{Hamana \&
  Mellier}{2001}]{Hamana2001}
Hamana T.,  Mellier Y.,  2001, \mn@doi [MNRAS]
  {10.1046/j.1365-8711.2001.04685.x}, 327, 169

\bibitem[\protect\citeauthoryear{Harris et~al.,}{Harris
  et~al.}{2020}]{Harris2020}
Harris C.~R.,  et~al., 2020, \mn@doi [Nature] {10.1038/s41586-020-2649-2}, 585,
  357

\bibitem[\protect\citeauthoryear{Hayashi et~al.,}{Hayashi
  et~al.}{2019}]{Hayashi2019}
Hayashi M.,  et~al., 2019, \mn@doi [PASJ] {10.1093/pasj/psz097}, 71, 1

\bibitem[\protect\citeauthoryear{Heymans et~al.,}{Heymans
  et~al.}{2012}]{Heymans2012}
Heymans C.,  et~al., 2012, \mn@doi [MNRAS] {10.1111/j.1365-2966.2012.21952.x},
  427, 146

\bibitem[\protect\citeauthoryear{Heymans et~al.,}{Heymans
  et~al.}{2013}]{Heymans2013}
Heymans C.,  et~al., 2013, \mn@doi [MNRAS] {10.1093/mnras/stt601}, 432, 2433

\bibitem[\protect\citeauthoryear{Higuchi \& Inoue}{Higuchi \&
  Inoue}{2018}]{Higuchi2018}
Higuchi Y.,  Inoue K.~T.,  2018, \mn@doi [MNRAS] {10.1093/mnras/sty205}, 476,
  359

\bibitem[\protect\citeauthoryear{Higuchi \& Inoue}{Higuchi \&
  Inoue}{2019}]{Higuchi2019}
Higuchi Y.,  Inoue K.~T.,  2019, \mn@doi [MNRAS] {10.1093/mnras/stz2150}, 488,
  5811

\bibitem[\protect\citeauthoryear{Higuchi \& Shirasaki}{Higuchi \&
  Shirasaki}{2016}]{Higuchi2016}
Higuchi Y.,  Shirasaki M.,  2016, \mn@doi [MNRAS] {10.1093/mnras/stw814}, 459,
  2762

\bibitem[\protect\citeauthoryear{Higuchi, Oguri  \& Hamana}{Higuchi
  et~al.}{2013}]{Higuchi2013}
Higuchi Y.,  Oguri M.,   Hamana T.,  2013, \mn@doi [MNRAS]
  {10.1093/mnras/stt521}, 432, 1021

\bibitem[\protect\citeauthoryear{Hikage et~al.,}{Hikage
  et~al.}{2019}]{Hikage2019}
Hikage C.,  et~al., 2019, \mn@doi [PASJ] {10.1093/pasj/psz010}, 71, 1

\bibitem[\protect\citeauthoryear{Hildebrandt et~al.,}{Hildebrandt
  et~al.}{2017}]{Hildebrandt2017a}
Hildebrandt H.,  et~al., 2017, \mn@doi [MNRAS] {10.1093/mnras/stw2805}, 465,
  1454

\bibitem[\protect\citeauthoryear{Hinshaw et~al.,}{Hinshaw
  et~al.}{2013}]{Hinshaw2013}
Hinshaw G.,  et~al., 2013, \mn@doi [ApJS] {10.1088/0067-0049/208/2/19}, 208, 19

\bibitem[\protect\citeauthoryear{Hunter}{Hunter}{2007}]{Hunter2007}
Hunter J.~D.,  2007, \mn@doi [Comput. Sci. Eng.] {10.1109/MCSE.2007.55}, 9, 90

\bibitem[\protect\citeauthoryear{Inoue \& Silk}{Inoue \&
  Silk}{2006}]{Inoue2006}
Inoue K.~T.,  Silk J.,  2006, \mn@doi [ApJ] {10.1086/505636}, 648, 23

\bibitem[\protect\citeauthoryear{Inoue \& Silk}{Inoue \&
  Silk}{2007}]{Inoue2007}
Inoue K.~T.,  Silk J.,  2007, \mn@doi [ApJ] {10.1086/517603}, 664, 650

\bibitem[\protect\citeauthoryear{Ivezi{\'{c}} et~al.,}{Ivezi{\'{c}}
  et~al.}{2019}]{Ivezic2019}
Ivezi{\'{c}} {\v{Z}}.,  et~al., 2019, \mn@doi [ApJ] {10.3847/1538-4357/ab042c},
  873, 111

\bibitem[\protect\citeauthoryear{Kawanomoto et~al.,}{Kawanomoto
  et~al.}{2018}]{Kawanomoto2018}
Kawanomoto S.,  et~al., 2018, \mn@doi [PASJ] {10.1093/pasj/psy056}, 70, 1

\bibitem[\protect\citeauthoryear{Kirshner, {Oemler, A.}, Schechter  \&
  Shectman}{Kirshner et~al.}{1981}]{Kirshner1981}
Kirshner R.~P.,  {Oemler, A.} J.,  Schechter P.~L.,   Shectman S.~A.,  1981,
  \mn@doi [ApJ] {10.1086/183623}, 248, L57

\bibitem[\protect\citeauthoryear{Komiyama et~al.,}{Komiyama
  et~al.}{2018}]{Komiyama2018}
Komiyama Y.,  et~al., 2018, \mn@doi [PASJ] {10.1093/pasj/psx069}, 70, 1

\bibitem[\protect\citeauthoryear{Lemaux, Tomczak, Lubin, Wu, Gal, Rumbaugh,
  Kocevski  \& Squires}{Lemaux et~al.}{2017}]{Lemaux2017}
Lemaux B.~C.,  Tomczak A.~R.,  Lubin L.~M.,  Wu P.-F.,  Gal R.~R.,  Rumbaugh
  N.,  Kocevski D.~D.,   Squires G.~K.,  2017, \mn@doi [MNRAS]
  {10.1093/mnras/stx1579}, 472, 419

\bibitem[\protect\citeauthoryear{Lewis, Challinor  \& Lasenby}{Lewis
  et~al.}{2000}]{Lewis2000}
Lewis A.,  Challinor A.,   Lasenby A.,  2000, \mn@doi [ApJ] {10.1086/309179},
  538, 473

\bibitem[\protect\citeauthoryear{Lietzen et~al.,}{Lietzen
  et~al.}{2016}]{Lietzen2016}
Lietzen H.,  et~al., 2016, \mn@doi [A{\&}A] {10.1051/0004-6361/201628261}, 588,
  L4

\bibitem[\protect\citeauthoryear{Lin et~al.,}{Lin et~al.}{2017}]{Lin2017a}
Lin Y.-T.,  et~al., 2017, \mn@doi [ApJ] {10.3847/1538-4357/aa9bf5}, 851, 139

\bibitem[\protect\citeauthoryear{Liske et~al.,}{Liske et~al.}{2015}]{Liske2015}
Liske J.,  et~al., 2015, \mn@doi [MNRAS] {10.1093/mnras/stv1436}, 452, 2087

\bibitem[\protect\citeauthoryear{Lubin, Brunner, Metzger, Postman  \&
  Oke}{Lubin et~al.}{2000}]{Lubin2000}
Lubin L.~M.,  Brunner R.,  Metzger M.~R.,  Postman M.,   Oke J.~B.,  2000,
  \mn@doi [ApJ] {10.1086/312518}, 531, L5

\bibitem[\protect\citeauthoryear{Lubin, Gal, Lemaux, Kocevski  \&
  Squires}{Lubin et~al.}{2009}]{Lubin2009}
Lubin L.~M.,  Gal R.~R.,  Lemaux B.~C.,  Kocevski D.~D.,   Squires G.~K.,
  2009, \mn@doi [AJ] {10.1088/0004-6256/137/6/4867}, 137, 4867

\bibitem[\protect\citeauthoryear{Mandelbaum, Seljak, Cool, Blanton, Hirata  \&
  Brinkmann}{Mandelbaum et~al.}{2006}]{Mandelbaum2006}
Mandelbaum R.,  Seljak U.,  Cool R.~J.,  Blanton M.,  Hirata C.~M.,   Brinkmann
  J.,  2006, \mn@doi [MNRAS] {10.1111/j.1365-2966.2006.10906.x}, 372, 758

\bibitem[\protect\citeauthoryear{Mandelbaum et~al.,}{Mandelbaum
  et~al.}{2018}]{Mandelbaum2018}
Mandelbaum R.,  et~al., 2018, \mn@doi [PASJ] {10.1093/pasj/psx130}, 70, 2

\bibitem[\protect\citeauthoryear{McKinney}{McKinney}{2010}]{McKinney2010}
McKinney W.,  2010, Proc. 9th Python Sci. Conf., 1697900, 51

\bibitem[\protect\citeauthoryear{Merloni et~al.,}{Merloni
  et~al.}{2012}]{Merloni2012}
Merloni A.,  et~al., 2012, arXiv e-prints, arXiv:1209.3114

\bibitem[\protect\citeauthoryear{Miyazaki et~al.,}{Miyazaki
  et~al.}{2018}]{Miyazaki2018}
Miyazaki S.,  et~al., 2018, \mn@doi [PASJ] {10.1093/pasj/psx063}, 70

\bibitem[\protect\citeauthoryear{Moutard, Sawicki, Arnouts, Golob, Coupon,
  Ilbert, Yang  \& Gwyn}{Moutard et~al.}{2020}]{Moutard2020}
Moutard T.,  Sawicki M.,  Arnouts S.,  Golob A.,  Coupon J.,  Ilbert O.,  Yang
  X.,   Gwyn S.,  2020, \mn@doi [MNRAS] {10.1093/mnras/staa706}, 494, 1894

\bibitem[\protect\citeauthoryear{Nicola et~al.,}{Nicola
  et~al.}{2020}]{Nicola2020}
Nicola A.,  et~al., 2020, \mn@doi [J. Cosmol. Astropart. Phys.]
  {10.1088/1475-7516/2020/03/044}, 2020, 044

\bibitem[\protect\citeauthoryear{Nishizawa et~al.,}{Nishizawa
  et~al.}{2018}]{Nishizawa2018}
Nishizawa A.~J.,  et~al., 2018, \mn@doi [PASJ] {10.1093/pasj/psx106}, 70, 1

\bibitem[\protect\citeauthoryear{Oguri}{Oguri}{2014}]{Oguri2014}
Oguri M.,  2014, \mn@doi [MNRAS] {10.1093/mnras/stu1446}, 444, 147

\bibitem[\protect\citeauthoryear{Oguri, Bayliss, Dahle, Sharon, Gladders,
  Natarajan, Hennawi  \& Koester}{Oguri et~al.}{2012}]{Oguri2012}
Oguri M.,  Bayliss M.~B.,  Dahle H.,  Sharon K.,  Gladders M.~D.,  Natarajan
  P.,  Hennawi J.~F.,   Koester B.~P.,  2012, \mn@doi [MNRAS]
  {10.1111/j.1365-2966.2011.20248.x}, 420, 3213

\bibitem[\protect\citeauthoryear{Oguri et~al.,}{Oguri et~al.}{2018}]{Oguri2018}
Oguri M.,  et~al., 2018, \mn@doi [PASJ] {10.1093/pasj/psx042}, 70, 1

\bibitem[\protect\citeauthoryear{Okabe, Takada, Umetsu, Futamase  \&
  Smith}{Okabe et~al.}{2010}]{Okabe2010}
Okabe N.,  Takada M.,  Umetsu K.,  Futamase T.,   Smith G.~P.,  2010, \mn@doi
  [PASJ] {10.1093/pasj/62.3.811}, 62, 811

\bibitem[\protect\citeauthoryear{Oke \& Gunn}{Oke \& Gunn}{1983}]{Oke1983}
Oke J.~B.,  Gunn J.~E.,  1983, \mn@doi [ApJ] {10.1086/160817}, 266, 713

\bibitem[\protect\citeauthoryear{Oort}{Oort}{1983}]{Oort1983}
Oort J.~H.,  1983, \mn@doi [ARA{\&}A] {10.1146/annurev.aa.21.090183.002105},
  21, 373

\bibitem[\protect\citeauthoryear{Park, Choi, Vogeley, {Gott III}  \&
  Blanton}{Park et~al.}{2007}]{Park2007}
Park C.,  Choi Y.,  Vogeley M.~S.,  {Gott III} J.~R.,   Blanton M.~R.,  2007,
  \mn@doi [ApJ] {10.1086/511059}, 658, 898

\bibitem[\protect\citeauthoryear{Paulino-Afonso, Sobral, Darvish, Ribeiro,
  Stroe, Best, Afonso  \& Matsuda}{Paulino-Afonso
  et~al.}{2018}]{Paulino-Afonso2018}
Paulino-Afonso A.,  Sobral D.,  Darvish B.,  Ribeiro B.,  Stroe A.,  Best P.,
  Afonso J.,   Matsuda Y.,  2018, \mn@doi [A{\&}A]
  {10.1051/0004-6361/201832688}, 620, A186

\bibitem[\protect\citeauthoryear{Peebles}{Peebles}{2001}]{Peebles2001}
Peebles P. J.~E.,  2001, \mn@doi [ApJ] {10.1086/322254}, 557, 495

\bibitem[\protect\citeauthoryear{Peng et~al.,}{Peng et~al.}{2010}]{Peng2010}
Peng Y.-j.,  et~al., 2010, \mn@doi [ApJ] {10.1088/0004-637X/721/1/193}, 721,
  193

\bibitem[\protect\citeauthoryear{Pisani, Sutter, Hamaus, Alizadeh, Biswas,
  Wandelt  \& Hirata}{Pisani et~al.}{2015}]{Pisani2015}
Pisani A.,  Sutter P.~M.,  Hamaus N.,  Alizadeh E.,  Biswas R.,  Wandelt B.~D.,
    Hirata C.~M.,  2015, \mn@doi [Phys. Rev. D] {10.1103/PhysRevD.92.083531},
  92

\bibitem[\protect\citeauthoryear{Regos \& Geller}{Regos \&
  Geller}{1991}]{Regos1991}
Regos E.,  Geller M.~J.,  1991, \mn@doi [ApJ] {10.1086/170332}, 377, 14

\bibitem[\protect\citeauthoryear{Robitaille et~al.,}{Robitaille
  et~al.}{2013}]{Robitaille2013}
Robitaille T.~P.,  et~al., 2013, \mn@doi [A{\&}A]
  {10.1051/0004-6361/201322068}, 558, A33

\bibitem[\protect\citeauthoryear{Rojas, Vogeley, Hoyle  \& Brinkmann}{Rojas
  et~al.}{2005}]{Rojas2005a}
Rojas R.~R.,  Vogeley M.~S.,  Hoyle F.,   Brinkmann J.,  2005, \mn@doi [ApJ]
  {10.1086/428476}, 624, 571

\bibitem[\protect\citeauthoryear{Sawicki et~al.,}{Sawicki
  et~al.}{2019}]{Sawicki2019}
Sawicki M.,  et~al., 2019, \mn@doi [MNRAS] {10.1093/mnras/stz2522}, 489, 5202

\bibitem[\protect\citeauthoryear{Scoville et~al.,}{Scoville
  et~al.}{2007}]{Scoville2007}
Scoville N.,  et~al., 2007, \mn@doi [ApJS] {10.1086/516585}, 172, 1

\bibitem[\protect\citeauthoryear{Shirasaki, Hamana  \& Yoshida}{Shirasaki
  et~al.}{2015}]{Shirasaki2015}
Shirasaki M.,  Hamana T.,   Yoshida N.,  2015, \mn@doi [MNRAS]
  {10.1093/mnras/stv1854}, 453, 3044

\bibitem[\protect\citeauthoryear{Shirasaki, Takada, Miyatake, Takahashi,
  Hamana, Nishimichi  \& Murata}{Shirasaki et~al.}{2017}]{Shirasaki2017}
Shirasaki M.,  Takada M.,  Miyatake H.,  Takahashi R.,  Hamana T.,  Nishimichi
  T.,   Murata R.,  2017, \mn@doi [MNRAS] {10.1093/mnras/stx1477}, 470, 3476

\bibitem[\protect\citeauthoryear{Shirasaki, Hamana, Takada, Takahashi  \&
  Miyatake}{Shirasaki et~al.}{2019}]{Shirasaki2019}
Shirasaki M.,  Hamana T.,  Takada M.,  Takahashi R.,   Miyatake H.,  2019,
  \mn@doi [MNRAS] {10.1093/mnras/stz791}, 486, 52

\bibitem[\protect\citeauthoryear{Sorrentino, Antonuccio-Delogu  \&
  Rifatto}{Sorrentino et~al.}{2006}]{Sorrentino2006}
Sorrentino G.,  Antonuccio-Delogu V.,   Rifatto A.,  2006, \mn@doi [A{\&}A]
  {10.1051/0004-6361:20065789}, 460, 673

\bibitem[\protect\citeauthoryear{Sousbie}{Sousbie}{2011}]{Sousbie2011}
Sousbie T.,  2011, \mn@doi [MNRAS] {10.1111/j.1365-2966.2011.18394.x}, 414, 350

\bibitem[\protect\citeauthoryear{Springel}{Springel}{2005}]{Springel2005a}
Springel V.,  2005, \mn@doi [MNRAS] {10.1111/j.1365-2966.2005.09655.x}, 364,
  1105

\bibitem[\protect\citeauthoryear{Steinhardt et~al.,}{Steinhardt
  et~al.}{2014}]{Steinhardt2014}
Steinhardt C.~L.,  et~al., 2014, \mn@doi [ApJ] {10.1088/2041-8205/791/2/L25},
  791, L25

\bibitem[\protect\citeauthoryear{Szapudi et~al.,}{Szapudi
  et~al.}{2015}]{Szapudi2015}
Szapudi I.,  et~al., 2015, \mn@doi [MNRAS] {10.1093/mnras/stv488}, 450, 288

\bibitem[\protect\citeauthoryear{{Tadaki}, {Iye}, {Fukumoto}, {Hayashi},
  {Rusu}, {Shimakawa}  \& {Tosaki}}{{Tadaki} et~al.}{2020}]{Tadaki2020}
{Tadaki} K.-i.,  {Iye} M.,  {Fukumoto} H.,  {Hayashi} M.,  {Rusu} C.~E.,
  {Shimakawa} R.,   {Tosaki} T.,  2020, arXiv e-prints, \href
  {https://ui.adsabs.harvard.edu/abs/2020arXiv200613544T} {arXiv:2006.13544}

\bibitem[\protect\citeauthoryear{Takada et~al.,}{Takada
  et~al.}{2014}]{Takada2014}
Takada M.,  et~al., 2014, \mn@doi [PASJ] {10.1093/pasj/pst019}, 66, R1

\bibitem[\protect\citeauthoryear{Takahashi, Hamana, Shirasaki, Namikawa,
  Nishimichi, Osato  \& Shiroyama}{Takahashi et~al.}{2017}]{Takahashi2017}
Takahashi R.,  Hamana T.,  Shirasaki M.,  Namikawa T.,  Nishimichi T.,  Osato
  K.,   Shiroyama K.,  2017, \mn@doi [ApJ] {10.3847/1538-4357/aa943d}, 850, 24

\bibitem[\protect\citeauthoryear{Tanaka}{Tanaka}{2015}]{Tanaka2015}
Tanaka M.,  2015, \mn@doi [ApJ] {10.1088/0004-637X/801/1/20}, 801, 20

\bibitem[\protect\citeauthoryear{Tanaka, Goto, Okamura, Shimasaku  \&
  Brinkmann}{Tanaka et~al.}{2004}]{Tanaka2004}
Tanaka M.,  Goto T.,  Okamura S.,  Shimasaku K.,   Brinkmann J.,  2004, \mn@doi
  [AJ] {10.1086/425529}, 128, 2677

\bibitem[\protect\citeauthoryear{Tanaka et~al.,}{Tanaka
  et~al.}{2018}]{Tanaka2018}
Tanaka M.,  et~al., 2018, \mn@doi [PASJ] {10.1093/pasj/psx077}, 70, 1

\bibitem[\protect\citeauthoryear{Taylor}{Taylor}{2005}]{Taylor2005}
Taylor M.~B.,  2005, Astron. Data Anal. Softw. Syst. XIV - ASP Conf. Ser., 347,
  29

\bibitem[\protect\citeauthoryear{Toshikawa et~al.,}{Toshikawa
  et~al.}{2018}]{Toshikawa2018}
Toshikawa J.,  et~al., 2018, \mn@doi [PASJ] {10.1093/pasj/psx102}, 70, 1

\bibitem[\protect\citeauthoryear{York et~al.,}{York et~al.}{2000}]{York2000}
York D.~G.,  et~al., 2000, \mn@doi [AJ] {10.1086/301513}, 120, 1579

\bibitem[\protect\citeauthoryear{de Jong, {Verdoes Kleijn}, Kuijken  \&
  Valentijn}{de~Jong et~al.}{2013}]{DeJong2013}
de Jong J. T.~A.,  {Verdoes Kleijn} G.~A.,  Kuijken K.~H.,   Valentijn E.~A.,
  2013, \mn@doi [Exp. Astron.] {10.1007/s10686-012-9306-1}, 35, 25

\bibitem[\protect\citeauthoryear{{de Jong} et~al.,}{{de Jong}
  et~al.}{2019}]{DeJong2019}
{de Jong} R.~S.,  et~al., 2019, \mn@doi [The Messenger]
  {10.18727/0722-6691/5117}, \href
  {https://ui.adsabs.harvard.edu/abs/2019Msngr.175....3D} {175, 3}

\bibitem[\protect\citeauthoryear{de
  Vaucouleurs}{de~Vaucouleurs}{1953}]{DeVaucouleurs1953}
de Vaucouleurs G.,  1953, \mn@doi [AJ] {10.1086/106805}, 58, 30

\makeatother
\end{thebibliography}


\appendix

\section{Density map catalogue}
\label{a1}

The projected density map catalogue at $z=$ 0.3 -- 1 is available as grid-point 
data through the HSC-SSP 
website\footnote{\url{https://hsc.mtk.nao.ac.jp/ssp/data-release/}}. 
We set an adequately small grid size of $\Delta=\sim1.5\times1.5$ square arcmin to the 
aperture size ($r=$ 10 or 30 arcmin) so that e.g., one can smooth and adjust the 
aperture size to the same comoving scale across the redshift bins if desired. Each 
pixel includes the number densities and the number excesses within each aperture 
at seven redshift slices: $z=$ [0.3:0.4), [0.4:0.5), [0.5:0.6), [0.6:0.7), 
[0.7:0.8), [0.8:0.9), and [0.9:1.0). The density information broken down to the 
narrower redshift range of $\Delta z=0.02$ is available as well. The accessible 
information is summarised in Table~\ref{tab1} where {\tt delta} is the number 
excess in variance, $(n_\mathrm{r}-n_\mathrm{r,mean})/n_\mathrm{r,mean}$. 
$n_\mathrm{r}$ is a number density within a radius of 10 arcmin or 30 arcmin, and 
$n_\mathrm{r,mean}$ is the mean value of $n_\mathrm{r}$ at each redshift in the 
whole survey area. 
The columns \#4 {\tt ix} and \#5 {\tt iy} are grid id which correspond to the 
grid size in each survey field summarised in Table~\ref{tab1}. They may be useful 
for reproducing the projected density map like Figure~\ref{fig6}-\ref{fig8}. 
The following script is an example to define a mesh grid of the W01 HectoMAP field 
for a Python visualisation tool: {\tt matplotlib} \citep{Hunter2007} with 
{\tt numpy} package \citep{Harris2020}, 

\begin{verbatim}
xmin, xmax = 224, 250
ymin, ymax = 42, 45
nx, ny = 754, 120
x0 = numpy.linspace(xmin, xmax, nx, endpoint=False)
y0 = numpy.linspace(ymin, ymax, ny, endpoint=False)
X, Y = numpy.meshgrid(x0, y0),
\end{verbatim}
and then users can associate the grid points with density values in the target 
redshift range through {\tt ix} and {\tt iy}. 

We should noted that, given the wide-field coverage, we have not conducted visual 
data inspections for individual regions. Thus, users must handle the catalogue 
with care at their own risk. For instance, when users make a plan of a 
spectroscopic follow-up observation towards their interesting fields based on the 
catalogue, we highly recommend to check the actual images taken from HSC around 
the targets as the minimum quality check through the user-friendly online 
visualisation tool called
{\tt hscMAP}\footnote{\url{https://hsc-release.mtk.nao.ac.jp/hscMap-pdr2/app/}}. 
Users may also want to check the more detailed image quality such as the seeing 
size and the image depth. In such a case, refer \#6 {\tt skymap\_id} to obtain the 
detailed information of the appropriate {\tt track} and {\tt patch}. The survey 
field of HSC-SSP is split to $\sim1.7\times1.7$ degree areas called {\tt tract}, 
and further divided into $\sim12\times12$ arcmin field called {\tt patch} 
\citep{Aihara2019}. Users can find such information for each {\tt track} and 
{\tt patch} stored in the HSC-SSP PDR2 
database\footnote{\url{https://hsc-release.mtk.nao.ac.jp/doc/index.php/sample-page/pdr2/}}.

\begin{table}
 \caption{Contents of the density map catalogue.}
 \label{tab1a}
 \begin{tabular}{rll}
  \hline
  \# & Name & Description \\
  \hline
  1  & {\tt ra}          & R.A. [degree] of the centre of aperture\\
  2  & {\tt dec}         & Dec. [degree] of the centre of aperture\\
  3  & {\tt field}       & Field id (see Table~\ref{tab1}) \\
  4  & {\tt ix}          & Pixel id in each field on the R.A. axis \\
  5  & {\tt iy}          & Pixel id in each field on the Dec. axis\\
  6  & {\tt skymap\_id}  & Nearest {\tt tract} and {\tt patch} in HSC-SSP \\
  7  & {\tt eff\_r10}    & Fraction of effective area in $r$=10 arcmin\\
  8  & {\tt eff\_r30}    & Fraction of effective area in $r$=30 arcmin\\
  9  & {\tt nd3\_r10}    & The number of $z$=[0.3:0.4) sources in $r$=10$^\prime$\\
 10  & {\tt nd3\_r30}    & The number of $z$=[0.3:0.4) sources in $r$=30$^\prime$\\
 11  & {\tt sgm3\_r10}   & Contrast of {\tt nc3\_r10} in standard deviation\\
 12  & {\tt sgm3\_r30}   & Contrast of {\tt nc3\_r30} in standard deviation\\
 13  & {\tt dlt3\_r10}   & Contrast of {\tt nc3\_r10} in variance\\
 14  & {\tt dlt3\_r30}   & Contrast of {\tt nc3\_r30} in variance\\
 15  & {\tt n3\_02\_r10} & The number of $z$=[0.30:0.32) sources in $r$=10$^\prime$\\
 16  & {\tt n3\_02\_r30} & The number of $z$=[0.30:0.32) sources in $r$=30$^\prime$\\
 17  & {\tt n3\_24\_r10} & The number of $z$=[0.32:0.34) sources in $r$=10$^\prime$\\
 18  & {\tt n3\_24\_r30} & The number of $z$=[0.32:0.34) sources in $r$=30$^\prime$\\
 19  & {\tt n3\_46\_r10} & The number of $z$=[0.34:0.36) sources in $r$=10$^\prime$\\
 20  & {\tt n3\_46\_r30} & The number of $z$=[0.34:0.36) sources in $r$=30$^\prime$\\
 21  & {\tt n3\_68\_r10} & The number of $z$=[0.36:0.38) sources in $r$=10$^\prime$\\
 22  & {\tt n3\_68\_r30} & The number of $z$=[0.36:0.38) sources in $r$=30$^\prime$\\
 23  & {\tt n3\_80\_r10} & The number of $z$=[0.38:0.40) sources in $r$=10$^\prime$\\
 24  & {\tt n3\_80\_r30} & The number of $z$=[0.38:0.40) sources in $r$=30$^\prime$\\
\multicolumn{2}{l}{25--40} & Same as \#9--24 but for $z$=[0.4:0.5)\\
\multicolumn{2}{l}{41--46} & Same as \#9--24 but for $z$=[0.5:0.6)\\
\multicolumn{2}{l}{47--62} & Same as \#9--24 but for $z$=[0.6:0.7)\\
\multicolumn{2}{l}{63--78} & Same as \#9--24 but for $z$=[0.7:0.8)\\
\multicolumn{2}{l}{79--94} & Same as \#9--24 but for $z$=[0.8:0.9)\\
\multicolumn{2}{l}{95--110} & Same as \#9--24 but for $z$=[0.9:1.0)\\
111  & {\tt eff\_z3}    & Same as \#7 but for $r$=10 cMpc at $z$=[0.3:0.4)\\
112  & {\tt eff\_z4}    & Same as \#7 but for $r$=10 cMpc at $z$=[0.4:0.5)\\
113  & {\tt eff\_z5}    & Same as \#7 but for $r$=10 cMpc at $z$=[0.5:0.6)\\
114  & {\tt eff\_z6}    & Same as \#7 but for $r$=10 cMpc at $z$=[0.6:0.7)\\
115  & {\tt eff\_z7}    & Same as \#7 but for $r$=10 cMpc at $z$=[0.7:0.8)\\
116  & {\tt eff\_z8}    & Same as \#7 but for $r$=10 cMpc at $z$=[0.8:0.9)\\
117  & {\tt eff\_z9}    & Same as \#7 but for $r$=10 cMpc at $z$=[0.9:1.0)\\
118  & {\tt sgm3\_c10}  & Same as \#11 but for $r$=10 cMpc at $z$=[0.3:0.4)\\
119  & {\tt dlt3\_c10}  & Same as \#13 but for $r$=10 cMpc at $z$=[0.3:0.4)\\
120  & {\tt sgm4\_c10}  & Same as \#11 but for $r$=10 cMpc at $z$=[0.4:0.5)\\
121  & {\tt dlt4\_c10}  & Same as \#13 but for $r$=10 cMpc at $z$=[0.4:0.5)\\
122  & {\tt sgm5\_c10}  & Same as \#11 but for $r$=10 cMpc at $z$=[0.5:0.6)\\
123  & {\tt dlt5\_c10}  & Same as \#13 but for $r$=10 cMpc at $z$=[0.5:0.6)\\
124  & {\tt sgm6\_c10}  & Same as \#11 but for $r$=10 cMpc at $z$=[0.6:0.7)\\
125  & {\tt dlt6\_c10}  & Same as \#13 but for $r$=10 cMpc at $z$=[0.6:0.7)\\
126  & {\tt sgm7\_c10}  & Same as \#11 but for $r$=10 cMpc at $z$=[0.7:0.8)\\
127  & {\tt dlt7\_c10}  & Same as \#13 but for $r$=10 cMpc at $z$=[0.7:0.8)\\
128  & {\tt sgm8\_c10}  & Same as \#11 but for $r$=10 cMpc at $z$=[0.8:0.9)\\
129  & {\tt dlt8\_c10}  & Same as \#13 but for $r$=10 cMpc at $z$=[0.8:0.9)\\
130  & {\tt sgm9\_c10}  & Same as \#11 but for $r$=10 cMpc at $z$=[0.9:1.0)\\
131  & {\tt dlt9\_c10}  & Same as \#13 but for $r$=10 cMpc at $z$=[0.9:1.0)\\
  \hline
 \end{tabular}
\end{table}

\section{Impact of redshift uncertainties}
\label{a2}

This paper discusses the mass contents of various environments traced by the 
projected density map based on the mock catalogue from the cosmological simulation. 
While we employ a sufficiently wide redshift bin of $\Delta z=0.1$ relative to 
typical photo-$z$ uncertainties ($\sigma=0.04$), 
ideally, we have to incorporate photo-$z$ errors and contamination effects from 
outliers into the analyses. We therefore carry out a quick test to see how 
photo-$z$ uncertainties makes an impact on our discussions in this paper by 
building photo-$z$ variations into redshifts of the mock samples. 

\begin{figure}
    \centering
	\includegraphics[width=.95\columnwidth]{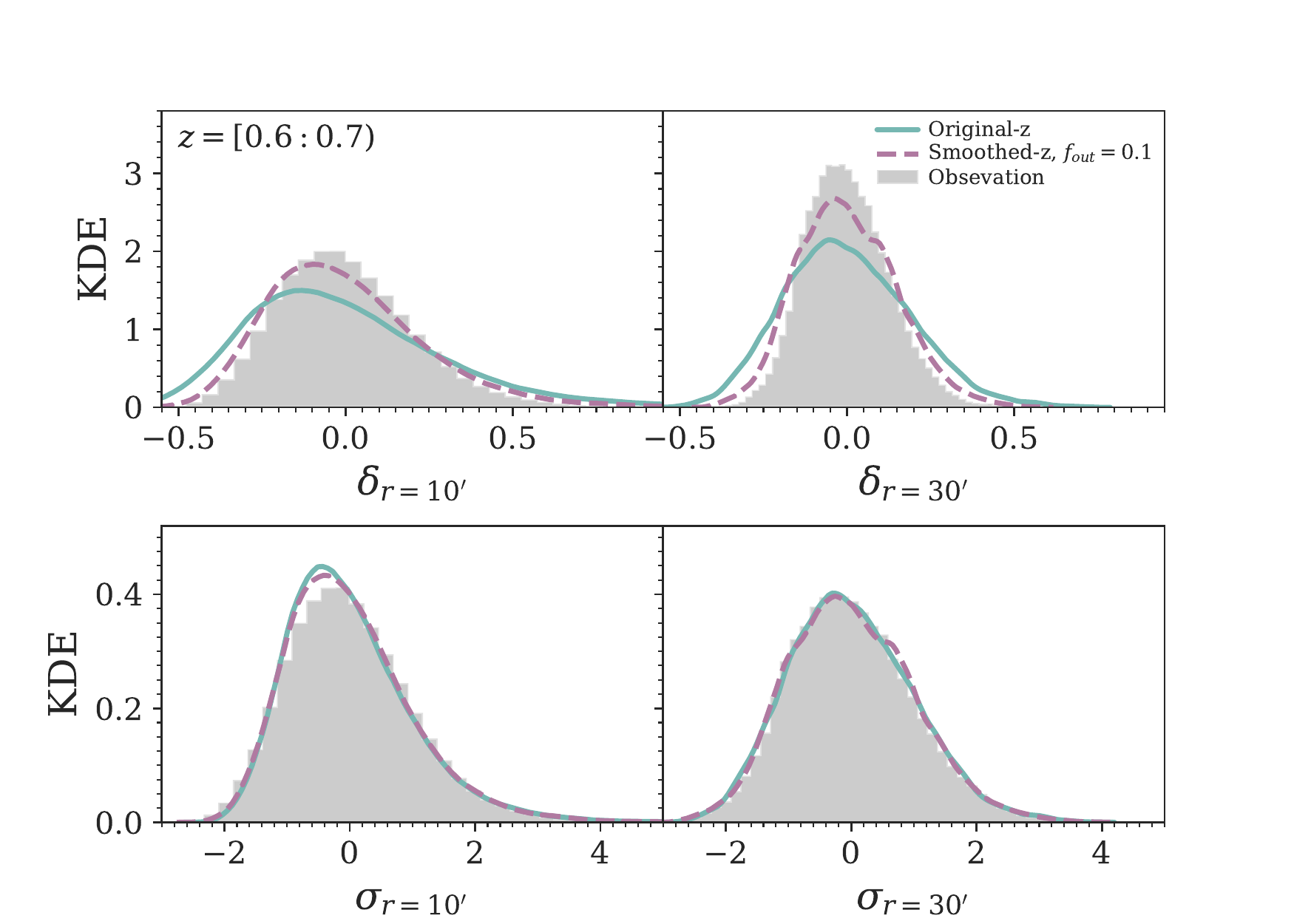}
    \caption{
    The upper and the lower panels respectively show variance and standard 
    deviation of number densities within (left) $r=10$ arcmin and (right) $r=30$ 
    arcmin apertures at $z=[0.6:0.7)$. 
    The grey filled histogram is the normalised density distribution from the 
    HSC-SSP PDR2 data (Fig.~\ref{fig4}). 
    The cyan curve (: original-$z$) and the purple dashed curves (: smoothed-$z$) 
    indicate those KDE from the mock catalogue with and without the redshift 
    variations, respectively (see text about the methodology). 
    }
    \label{fig1a}
\end{figure}

In order to test the impacts of photo-$z$ errors on our analyses, we choose the 
intermediate redshift bin $z=[0.6:0.7)$ out of seven redshift slices in $z=0.3$ -- 
1 that we used in this paper, which allows us to add irrelevant sources from 
distant foreground ($z=0.3$ -- 0.5) and background ($z=0.8$ -- 1) redshifts into 
the sample.
Firstly, we convolve redshifts of all mock sources at $z=0.3$ -- 1 with a 
Gaussian filter with $\sigma=0.04$ according to the typical redshift dispersion of 
photometric redshifts from {\tt Mizuki} for $i$-band ($i<23$) selected sources at 
$z<1$ \citep{Tanaka2018}.
After that, we measure the number densities within $r=10$ 
or $r=30$ arcmin apertures at $z=[0.6:0.7)$ in the same way as for the HSC sample 
and the mock data (\S\ref{s32} and \S\ref{s41}). In this process, we mix the 
randomly selected outliers in the foreground ($z=0.3$ -- 0.5) or the background 
($z=0.8$ -- 1) in with the sample at $z=[0.6:0.7)$. The number of outliers are 
arranged to 10 per cent the outlier rate, 
$f_{out}\equiv N_\mathrm{outlier}/N_\mathrm{total}=0.1$, which is consistent with 
that of our sample \citep{Tanaka2018}. 
It should be noted that our simplified test would not produce an exact photo-$z$ 
error although this tells how irrelevant sources affect the density estimation: 
photometric redshift errors of galaxies are not random distribution, rather, the 
errors should be biased to certain redshift ranges depending on those SEDs and 
redshifts.

\begin{figure}
    \centering
	\includegraphics[width=.95\columnwidth]{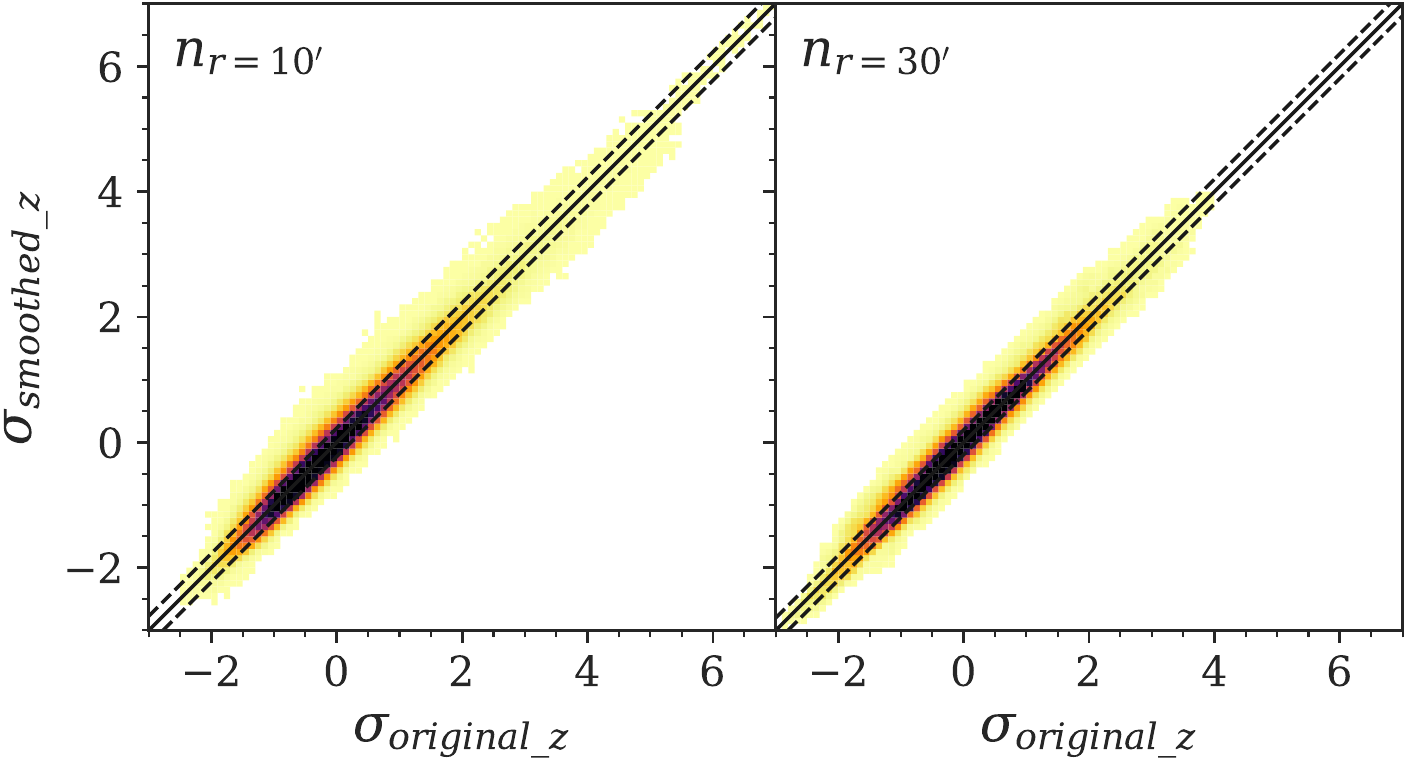}
    \caption{
    The number densities within $r=10$ arcmin (on the left) and $r=30$ arcmin (on 
    the right) apertures in the standard deviation inclusive photo-$z$ 
    uncertainties($\sigma_\mathrm{original z}$), with respect to those inferred 
    from the original mock sample ($\sigma_\mathrm{smoothed z}$). 
    The solid and dashed diagonal lines indicate a one-to-one relation and 
    $\pm1\sigma$ deviations of the density contrast between with and without 
    installing photo-$z$ errors.
    }
    \label{fig2a}
\end{figure}

\begin{figure}
    \centering
	\includegraphics[width=.9\columnwidth]{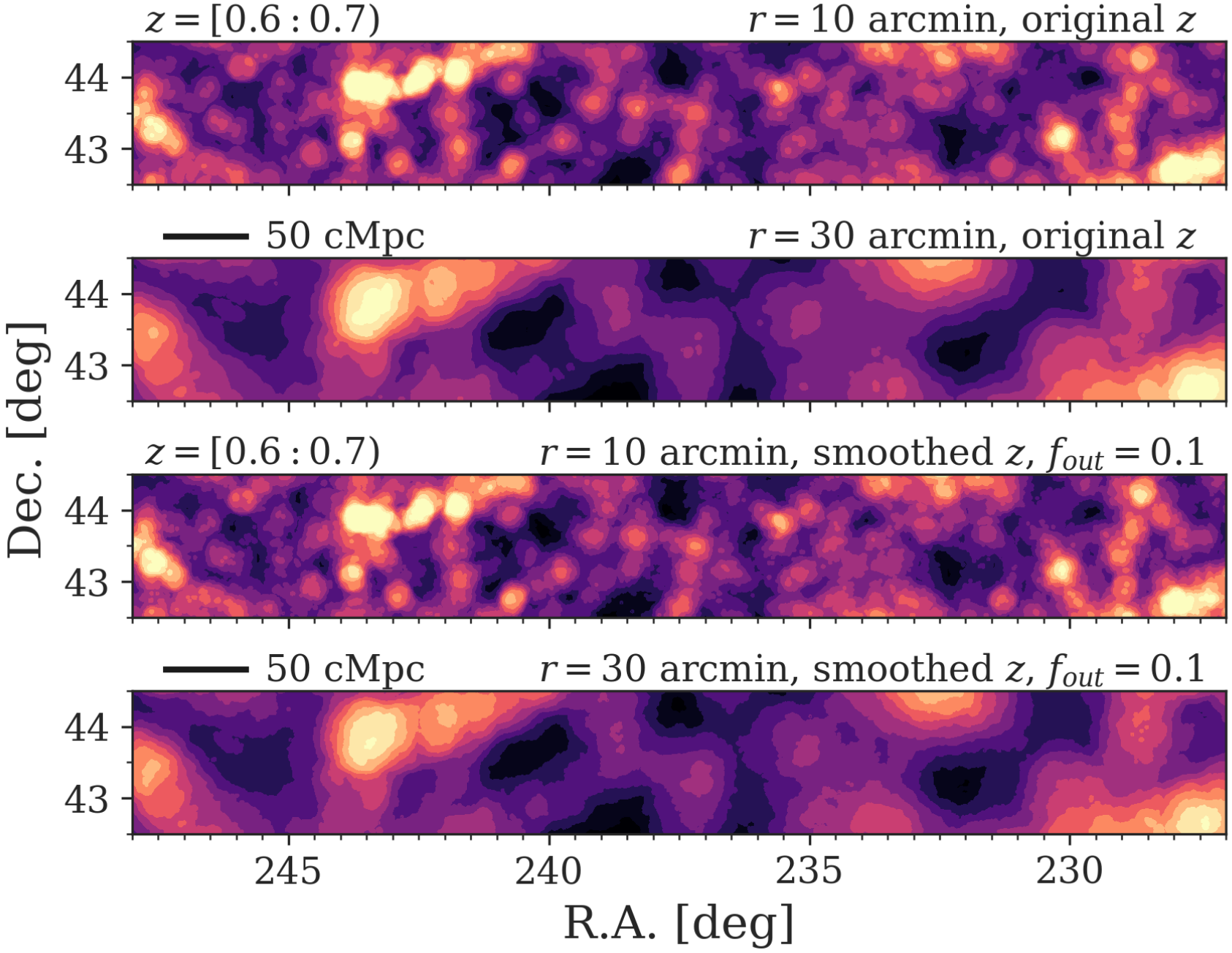}
    \caption{
    Example cutouts from the mock density maps at $z=[0.6:0.7)$ in W01 HectoMAP. 
    From top to bottom, the density maps within $r=10$ and 30 arcmin apertures 
    based on the original mock catalogue, and those within $r=10$ and 30 arcmin 
    apertures from the mock sample that incorporates photo-$z$ uncertainties. 
    }
    \label{fig3a}
\end{figure}

\begin{figure}
    \centering
	\includegraphics[width=.95\columnwidth]{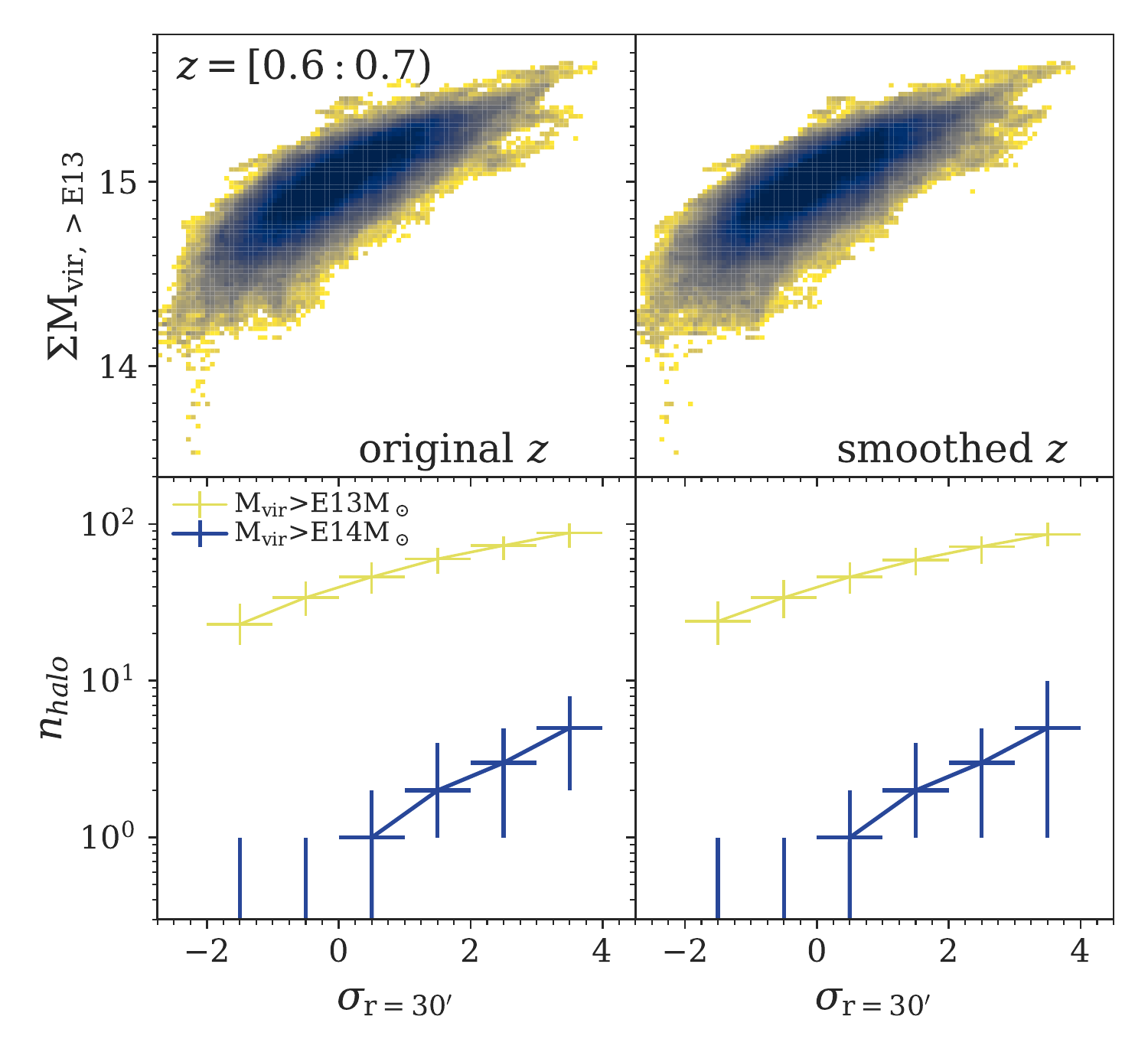}
    \caption{
    Same as Figure~\ref{fig13}, but for the density distributions at $z=[0.6:0.7)$ 
    without and with consideration of photo-$z$ uncertainties on the left and 
    right sides, respectively. 
    }
    \label{fig4a}
\end{figure}

The resultant density distribution in the variance and the standard deviation are 
presented in Figure~\ref{fig1a}. Those in consideration of photometric redshift 
errors show a broad agreement with the density distributions of the observed data, 
suggesting that the assumption of the large-scale galaxy bias 
\citep[eq.~4.11 and 4.12]{Nicola2020} in producing $i$-band selected mock sources 
works for our sample. Residual margin seen in the variances between the observation and the mock data (Fig.~\ref{fig1a}) can be minimised by assuming the outlier 
rate of $\sim0.2$, which suggests that our sample may contain more contaminant 
foreground and/or background sources. 
Photo-$z$ uncertainties scatter the density contrast of 
original spatial distributions, as shown in Figure~\ref{fig2a} and make the 
density variance spikier. Figure~\ref{fig1a} and \ref{fig2a} also argue that original mock density 
distribution out of consideration of photo-$z$ errors can be used to test 
systematic trends as long as we employ the standard deviation as a tracer of the 
density contrast. Such a consensus of the density contrasts between with and 
without photo-$z$ uncertainties can also be recognised in the mock projected 
density maps shown in Figure~\ref{fig3a}. As a result, photometric redshift 
errors have a little influence on the discussion, as seen in Figure~\ref{fig4a}.

\section{Online materials}
\label{a3}

\begin{landscape}
\begin{figure}
    \centering
	\includegraphics[width=1.0\linewidth]{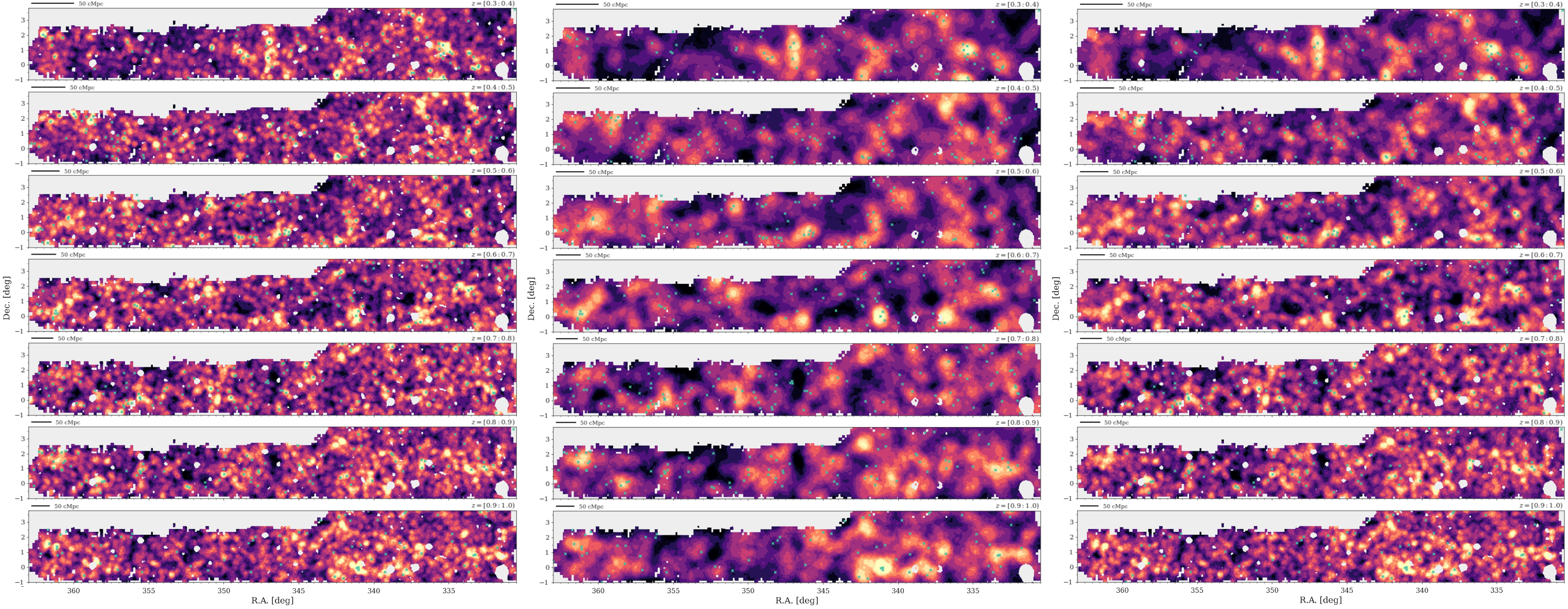}
    \caption{Same as Fig.~\ref{fig6}--~\ref{fig8} but for W05 (VVDS).}
    \label{fig1add}
\end{figure}
\end{landscape}

\begin{landscape}
\begin{figure}
    \centering
	\includegraphics[width=1.0\linewidth]{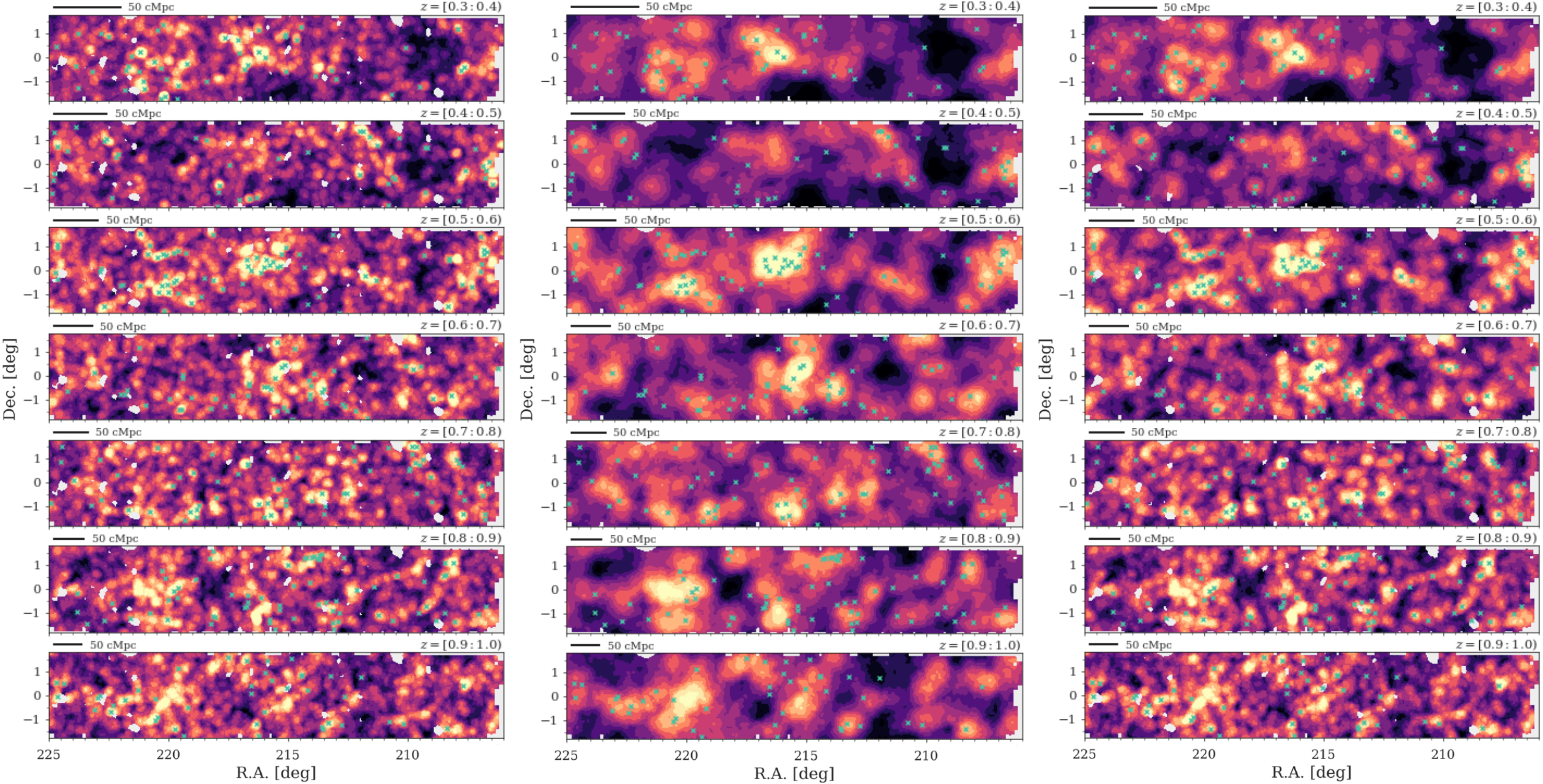}
    \caption{Same as Fig.~\ref{fig6}--~\ref{fig8} but for W04 (GAMA15H).}
    \label{fig2add}
\end{figure}
\end{landscape}

\begin{landscape}
\begin{figure}
    \centering
	\includegraphics[width=1.0\linewidth]{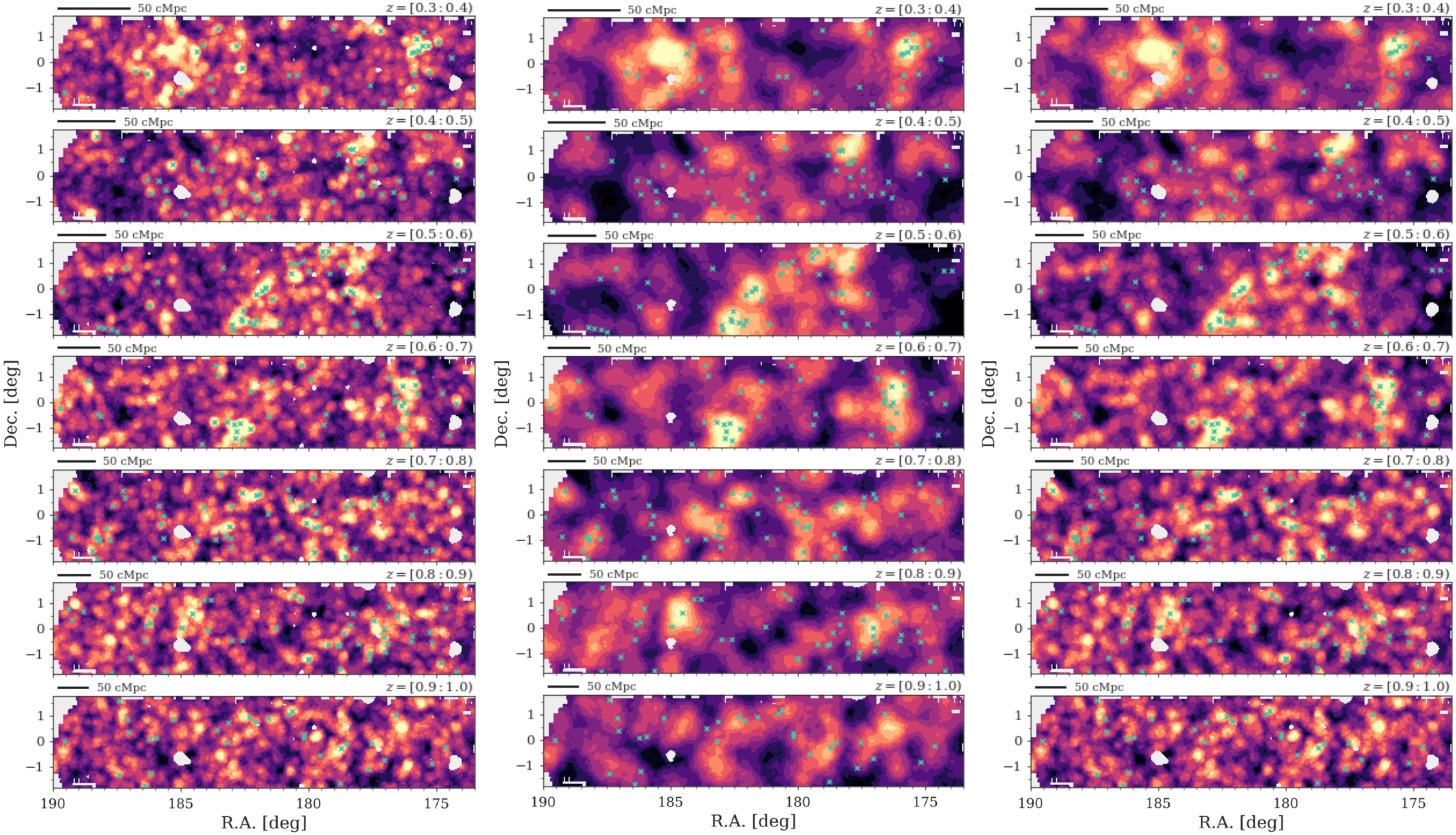}
    \caption{Same as Fig.~\ref{fig6}--~\ref{fig8} but for W04 (WIDE12H).}
    \label{fig3add}
\end{figure}
\end{landscape}

\begin{landscape}
\begin{figure}
    \centering
	\includegraphics[width=1.0\linewidth]{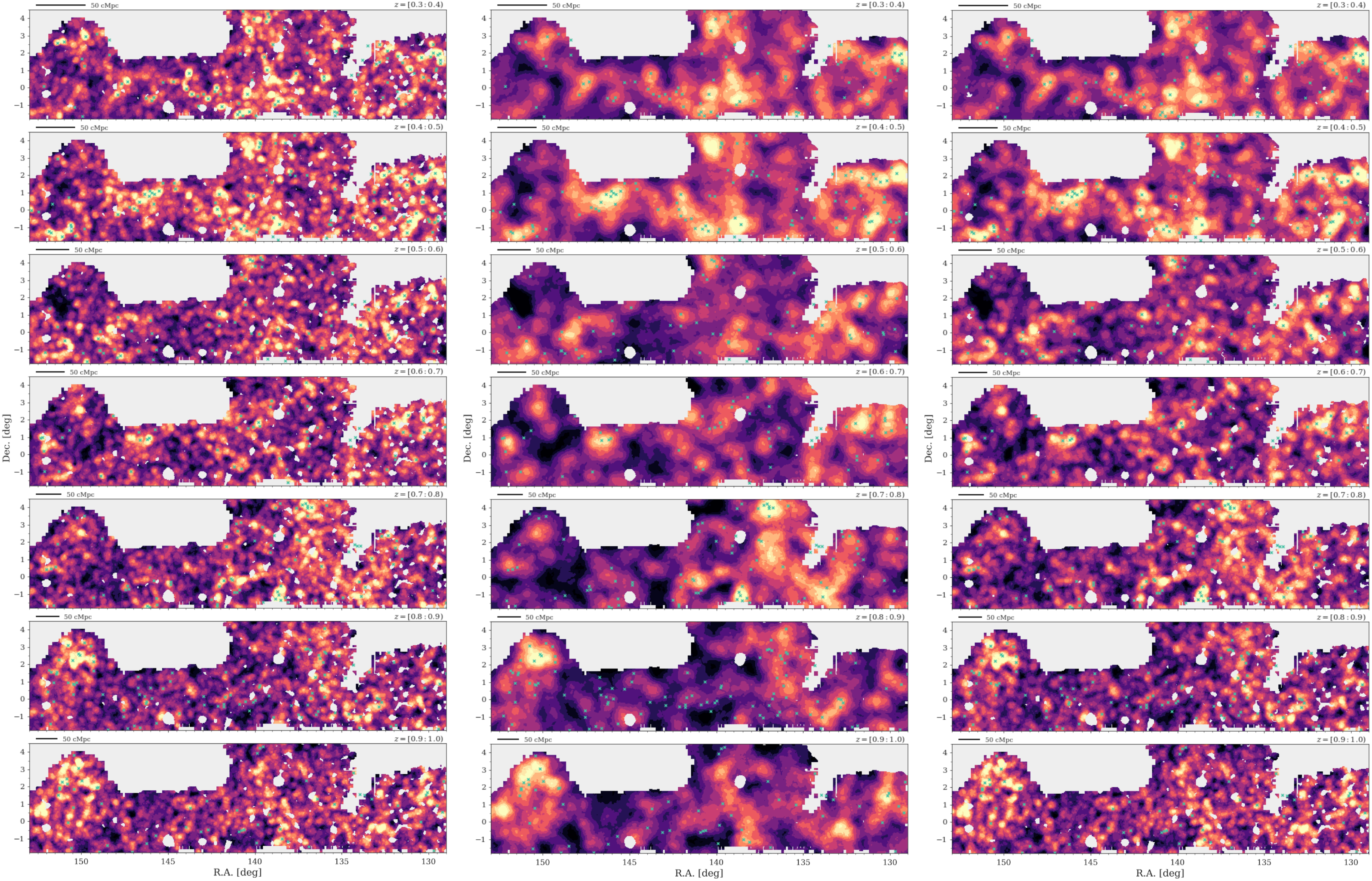}
    \caption{Same as Fig.~\ref{fig6}--~\ref{fig8} but for W03 (GAMA09H).}
    \label{fig4add}
\end{figure}
\end{landscape}


\bsp	
\label{lastpage}
\end{document}